\documentclass[prb,twocolumn,longbibliography,amsmath,superscriptaddress,amssymb]{revtex4-1}

\usepackage{graphicx}
\usepackage{dcolumn}
\usepackage{bm}
\usepackage{amssymb}
\usepackage{amsmath}
\usepackage{color}
\usepackage[dvipsnames]{xcolor}
\usepackage{placeins}
\usepackage{epstopdf}

\usepackage[normalem]{ulem}

\usepackage[linkcolor=blue,urlcolor=blue,citecolor=blue,colorlinks=true]{hyperref}

\newcommand{\spr}{\shortparallel}
\newcommand{\rtm}[1]{\mathrm{#1}}

\begin{document}

\title{\textit{Ab initio} ${\mathbf k}\cdot{\mathbf p}$ theory of spin-momentum locking: Application to topological surface states}

\author{I. A. Nechaev}
\affiliation{Department of Electricity and Electronics, FCT-ZTF, UPV-EHU, 48080 Bilbao, Spain}

\author{E. E. Krasovskii}
\affiliation{Donostia International Physics Center (DIPC), Paseo Manuel de Lardizabal 4, 20018 Donostia/San Sebasti\'{a}n,  Basque Country, Spain}
\affiliation{Departamento de F\'{i}sica de Materiales, Facultad de Ciencias Qu\'{i}micas, Universidad del Pais Vasco/Euskal Herriko Unibertsitatea, Apdo. 1072, 20080 Donostia/San Sebasti\'{a}n, Basque Country, Spain}
\affiliation{IKERBASQUE, Basque Foundation for Science, 48013 Bilbao, Basque Country, Spain}

\date{\today}

\begin{abstract}
Based on \textit{ab initio} relativistic ${\mathbf k}\cdot{\mathbf p}$ theory, we derive an effective two-band model for surface states of three-dimensional topological insulators up to seventh order in $\mathbf{k}$. It provides a  comprehensive description of the surface spin structure characterized by a non-orthogonality between momentum and spin. We show that the oscillation of the non-orthogonality with the polar angle of $\mathbf{k}$ with a $\pi/3$ periodicity can be seen as due to effective six-fold symmetric spin-orbit magnetic fields with a quintuple and septuple winding of the field vectors per single rotation of \textbf{k}. Owing to the dominant effect of the classical Rashba field, there remains a single-winding helical spin structure but with a periodic few-degree deviation from the orthogonal locking between momentum and spin.
\end{abstract}

\maketitle

\section{Introduction}

Over the last decade, the effective model Hamiltonian for topological surface states developed in Refs.~[\onlinecite{Zhang_NATPHYS_2009}] and~[\onlinecite{Liu_PRB_2010}] is commonly accepted as a tool to include in a simple manner their remarkable features: the linear energy-momentum dispersion and helical in-plane spin structure. This model has been applied to a variety of topologically non-trivial materials in the spirit of the classical Rashba model, which, since the seminal paper by LaShell et al.~[\onlinecite{LaShell_PRL_1996}], has been used to fit the two-dimensional (2D) spin-orbit-split states at trivial surfaces. The spin-orbit splitting $k_{\pm}$-linear term is the same for trivial and for topological surface states, and it yields an orthogonal spin-momentum locking commonly considered a hallmark of a strong spin-orbit interaction (SOI).

The simplified model of Refs.~[\onlinecite{Zhang_NATPHYS_2009}] and [\onlinecite{Liu_PRB_2010}] needs to be extended in order to describe the non-orthogonality between spin and momentum in realistic systems, as, e.g., observed in photoemission from Bi$_2$Se$_3$~[\onlinecite{Wang_PRL_2011}]. While it naturally arises in \textit{ab initio} calculations, in $\mathbf{k}\cdot\mathbf{p}$ theory, in order to yield a deviation from orthogonality, an effective Hamiltonian must include higher-order in $\mathbf{k}$ spin-orbit terms. In Ref.~[\onlinecite{Basak_PRB_2011}], for structures with the $C_{3v}$ crystal symmetry and time-reversal symmetry it was suggested to include a $k_{\pm}^5$ term to allow for the non-orthogonality, and a minimal fifth-order $\mathbf{k}\cdot\mathbf{p}$ model was applied to Bi$_2$Te$_3$. Following Ref.~[\onlinecite{Basak_PRB_2011}], in Ref.~[\onlinecite{Hopfner_PRL_2012}] this model was also used to analyze the spin structure of the Au/Ge(111) surface state, which is rather far from being Rashba-like. Since the fifth-order Hamiltonian was constructed based on symmetry arguments rather than derived directly from \textit{ab initio} spinor wave functions, its parameters were found by fitting to the \textit{ab initio} band structure, which is an approximate procedure sensitive to the choice of the energy interval of interest and to the order of the $\mathbf{k}\cdot\mathbf{p}$ expansion: with each successive order the complexity and ambiguity grow, and the parameters become increasingly less physically meaningful.

Here, we study the angle between the spin and momentum in the surface states of Bi$_2$Se$_3$ and Bi$_2$Te$_2$Se within our \textit{ab initio} relativistic $\mathbf{k}\cdot\mathbf{p}$ approach introduced in Refs.~[\onlinecite{Nechaev_PRBR_2016, Nechaev_PRB_2018, Nechaev_PRB_2019}]. This approach has been successfully applied to different materials~[\onlinecite{Nechaev_SciRep_2017, Nechaev_PRB_2019, Susanne2019, Usachov_arXiv_2020}] and established as a reliable theoretical tool for deriving few-band $\mathbf{k}\cdot\mathbf{p}$ Hamiltonians capable of comprehensive description of the surface spin structure. We take Bi$_2$Se$_3$ and Bi$_2$Te$_2$Se as vivid examples of the topological insulators (TIs) with a rather wide absolute bulk band gap bridged by the partly occupied topological surface state and a local projected gap well above the Fermi level hosting the so-called ``second topological surface state''~[\onlinecite{Niesner_PRB_2012, Sobota_PRL_2013, Niesner_JESRP_2014, Datzer_PRB_2017, Aguilera_PRB_2019}]. The wide gap is favorable for minimizing the effect of the proximity of bulk states on the surface-state spin structure. The presence of the second surface state makes it possible to derive a \textit{two-band seventh-order} Hamiltonian by applying the L\"{o}wdin partitioning to a \textit{four-band third-order} Hamiltonian generated for the two surface states within our \textit{ab initio} approach.

For Bi$_2$Se$_3$ and Bi$_2$Te$_2$Se, we derive the seventh-order Hamiltonian allowing for the non-orthogonal locking between momentum and spin. We show that there is an oscillation of the spin around the momentum-perpendicular direction with a $\pi/3$ periodicity as a function of the polar angle of $\mathbf{k}$ due to the presence of $\mathbf{k}$-dependent effective spin-orbit magnetic fields of the six-fold symmetry. The effective Hamiltonian facilitates the inclusion of the non-orthogonality in the description of spin-related properties of the TI surfaces and their interpretation within $\mathbf{k}\cdot\mathbf{p}$ theory. Thus, our study can also be considered as an \textit{ab initio} substantiation of the fifth-order Hamiltonian proposed in Ref.~[\onlinecite{Wang_PRL_2011}], based on an unambiguous algorithm for its parameters.

\section{Computational details}

The \textit{ab initio} band structure is obtained with the extended linear augmented plane waves method~[\onlinecite{Krasovskii_PRB_1997}] using the full potential scheme of Ref.~[\onlinecite{Krasovskii_PRB_1999}] within the local density approximation (LDA). The spin-orbit interaction was treated as a second variation~[\onlinecite{Koelling_1977}]. The surfaces of the TIs are simulated by bulk-truncated centrosymmetric six-QL (quintuple layer) slabs of space group $P\bar{3}m1$ (no.~164). The experimental crystal lattice parameters were taken from Ref.~[\onlinecite{Wyckoff_RWG}]. In the case of Bi$_2$Te$_2$Se, the experimental atomic positions of Ref.~[\onlinecite{Wyckoff_RWG}] were used, while for Bi$_2$Se$_3$ we took the LDA relaxed atomic positions of Ref.~[\onlinecite{Nechaev_PRB_2013_BISE}].

\section{Spin-momentum locking angle}

\begin{figure}[tbp]
\centering
\includegraphics[width=\columnwidth]{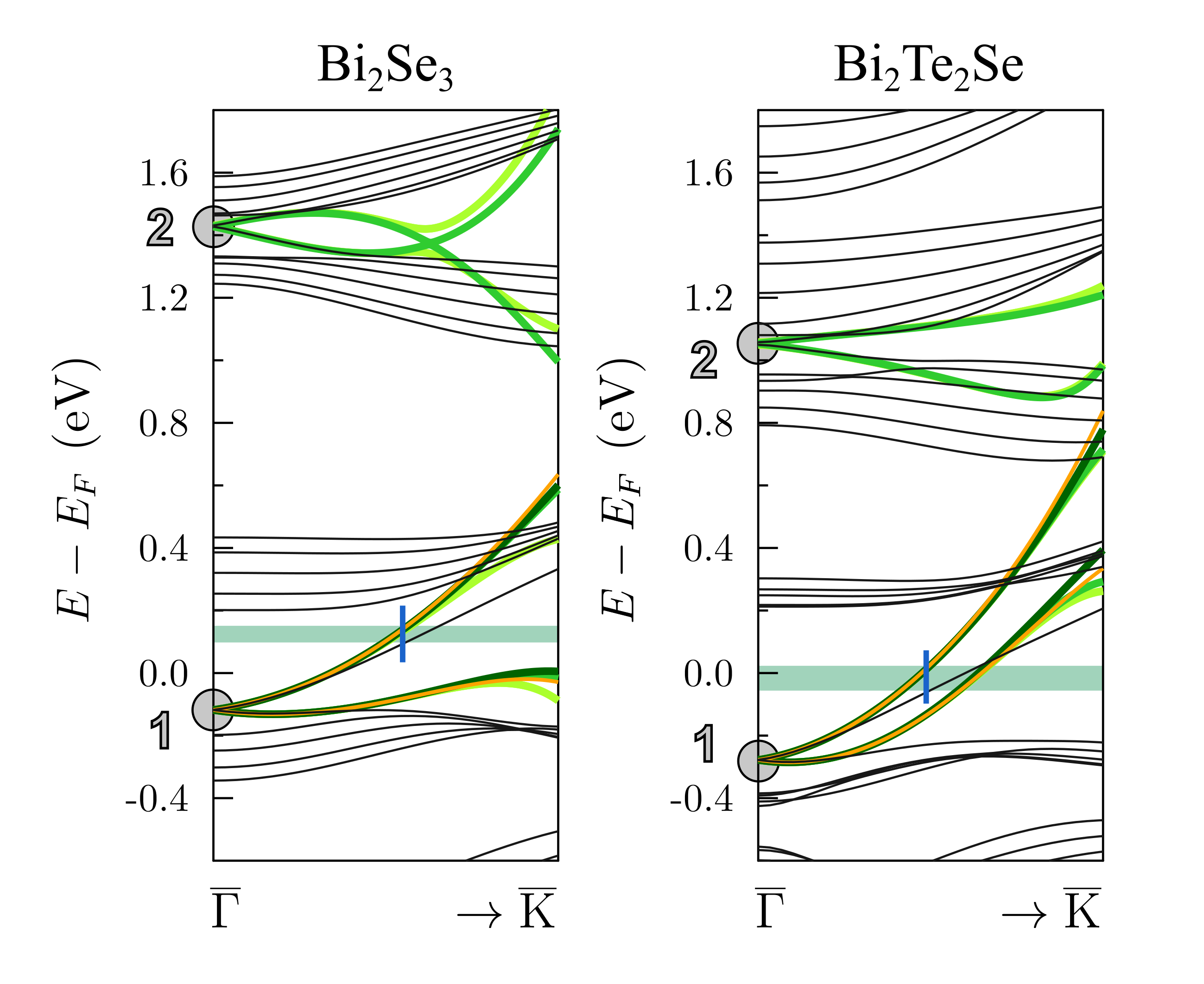}
\caption{Band structure of the surface of Bi$_2$Se$_3$ and Bi$_2$Te$_2$Se along $\bar{\Gamma}$-$\bar{K}$ by the full LDA Hamiltonian $H^{\rtm{LDA}}_{{\mathbf k}}$ (black lines), the $8\times8$ Hamiltonian~(\ref{HamTwoSurfaces}) (light green lines), the $4\times4$ Hamiltonian~(\ref{TWOSSHam}) (green lines), and the $2\times2$ Hamiltonian~(\ref{UPTO7}) with two different values of $\widetilde{\gamma}^{(7)}$ (dark green and orange lines, see text). The upper (lower) border of the horizontal stripes corresponds to the energy at which the model (\textit{ab initio}) CECs shown in Fig.~\ref{fig2} are calculated. The $\mathbf{k}$-points on these CECs along $\bar{\Gamma}$-$\bar{K}$ are marked by vertical blue lines.}
\label{fig1}
\end{figure}

\begin{figure*}[tbp]
\centering
\includegraphics[width=\textwidth]{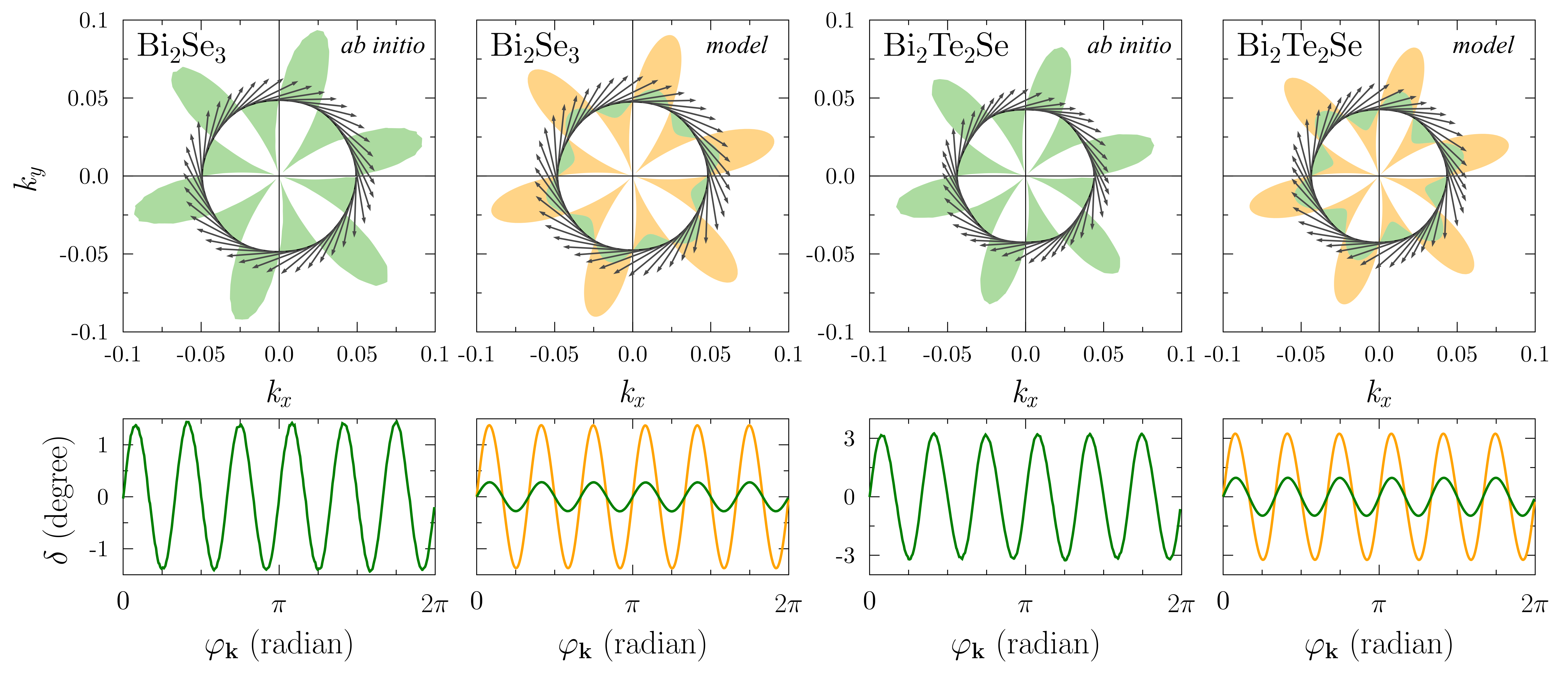}
\caption{Spin-resolved constant energy contours (upper panels) and the non-orthogonality $\delta$ as a function of the polar angle $\varphi_{\mathbf{k}}$ (lower panels) for Bi$_2$Se$_3$ and Bi$_2$Te$_2$Se. In the upper panels, the colored areas highlight the deviation of the in-plane spin direction from the classical Rashba in-plane spin at a given $\mathbf{k}$-point in the contour: the border of the areas is given by the length of the relevant $\mathbf{k}$ plus a scaled $\mathbf{k}$-projection of the in-plane spin: $k+R\sin\delta$, $R$ being the scaling factor. The green areas and lines correspond to the LDA and ${\mathbf k}\cdot{\mathbf p}$ calculations, while the orange ones -- to the ${\mathbf k}\cdot{\mathbf p}$ calculations with the magnitude of $\gamma^{(5)}$ manually increased by a factor of 4.5 for Bi$_2$Se$_3$ and of 3.8 for Bi$_2$Te$_2$Se, see text. The contours are calculated at the energies marked in Fig.~\ref{fig1}.}
\label{fig2}
\end{figure*}

Figure~\ref{fig1} shows the calculated LDA band structure of the Bi$_2$Se$_3$ and Bi$_2$Te$_2$Se surfaces along $\bar{\Gamma}$-$\bar{K}$. Two topological surface states are clearly identified in the spectra of both TIs. The two states are numbered $n=1$ and 2 in order of increasing energy. The spin-resolved constant energy contours (CECs) at $\sim0.2$~eV above the Dirac points of the low-energy surface states ($n=1$) are shown in Fig.~\ref{fig2} together with the respective angles of deviation from the orthogonal spin-momentum coupling $\delta$. As a function of the polar angle $\varphi_{{\mathbf k}}$ of the momentum $\mathbf{k}$, the deviation angle demonstrates an oscillating behavior with a $\pi/3$ periodicity and an amplitude close to 1.5$^{\circ}$ for Bi$_2$Se$_3$, which is in good agreement with the experiment of Ref.~[\onlinecite{Wang_PRL_2011}], and about 3.0$^{\circ}$ for Bi$_2$Te$_2$Se.

We start with a ${\mathbf k}\cdot{\mathbf p}$ model comprising both Dirac surface states $n=1$ and 2. Being eigenfunctions of a centrosymmetric slab Hamiltonian at $\bar{\Gamma}$, these states form four Kramers-degenerate pairs with spinor wave functions $\Psi_{m\mu}$, which we group into two twin pairs with two members, $m = 2n-1$ and $2n$. Here, $\mu=\uparrow$ or $\downarrow$ indicates the sign $+$ or $-$ of the expectation value $\langle\Psi_{m\mu}| \widehat{J}_{z} |\Psi_{m'\mu'}\rangle_{\tau}=\langle J_z\rangle_{m\mu}\delta_{mm'}\delta_{\mu\mu'}$ of the $z$ projection of the total angular momentum $\widehat{\mathbf{J}}$ at the (symmetry equivalent) atomic sites of type $\tau$, which has the largest weight $\langle\Psi_{m\mu}|\Psi_{m\mu}\rangle_{\tau}$, see Ref.~[\onlinecite{Nechaev_PRBR_2016}]. The integration is over the muffin-tin spheres of this type, and the positive value is $\langle J_z\rangle_{m\uparrow} = -\langle J_z\rangle_{m\downarrow}$. With this basis set, we first microscopically derive an eight-band $\mathbf{k}\cdot\mathbf{p}$ Hamiltonian $H_{\rtm{\mathbf{kp}}}$ from an \textit{ab initio} relativistic ${\mathbf k}\cdot{\mathbf p}$ perturbation expansion around the $\Gamma$ point. The expansion is carried out up to the third order in \textbf{k} by applying  the L\"{o}wdin partitioning~[\onlinecite{Leowdin_JCP_1951, Schrieffer_PR_1966, Winkler_KP}] to the original Hilbert space of the $\Gamma$-projected LDA Hamiltonian $H^{\rtm{LDA}}_{{\mathbf k}}$, see Appendix~\ref{A_expansion}.

Because $\Psi_{m\mu}$ are slab eigenfunctions representing the surface states, each twin-pair is characterized by two doubly degenerate slab levels $E_{2n-1}$ and $E_{2n}$ separated by $\Delta_n = E_{2n}-E_{2n-1}$ of a few meV due to the bonding-antibonding interaction. Since the $\bar{\Gamma}$ point is a TRIM (time reversal invariant momentum), the spinors $\Psi_{m\mu}$ are also parity eigenfunctions, and the two pairs for a given $n=1$ or 2 have different parity. We now transfer to a new basis $|\Phi^{\pm}_{n\mu}\rangle =\frac{1}{\sqrt{2}} \left[|\Psi_{2n-1\mu}\rangle \pm |\Psi_{2n\mu}\rangle\right]$, where the new basis functions $|\Phi^{\pm}_{n\mu}\rangle$ are no longer parity eigenfunctions but are localized at one of the two surfaces of the 6QL slab, ``$+$'' or ``$-$''. In this surface-resolved basis, the original $8\times8$ Hamiltonian reads
\begin{equation}\label{HamTwoSurfaces}
H_{\rtm{\mathbf{kp}}}\longrightarrow H^{\rtm{Film}}_{\rtm{\mathbf{kp}}}=\left(
\begin{array}{cc}
H_{\rtm{Surf}}^{+} & H_{\rtm{int}}  \\
H^{\dag}_{\rtm{int}} & H_{\rtm{Surf}}^{-}
\end{array}
\right).
\end{equation}
In Fig.~\ref{fig1}, the bands obtained by diagonalizing this Hamiltonian are shown by light green lines for both TIs.

Further we neglect the coupling of the surfaces due to the overlap between the $+$ and $-$ new basis functions, $H_{\rtm{int}}\rightarrow0$, and in the following we will consider only the $-$ surface, so we omit the superscript $-$. In a compact form, the resulting $4\times4$ Hamiltonian, which is just the term $H_{\rtm{Surf}}^{-}$ of the Hamiltonian~(\ref{HamTwoSurfaces}), reads
\begin{equation}\label{TWOSSHam}
H_{\rtm{\mathbf{kp}}}^{4\times4}=\left(
\begin{array}{cc}
E_1+H_1+H^{\rtm{R}}_1      & H_0 +\widetilde{H} \\
H_0 +\widetilde{H}^{\dag}  & E_2+H_2+H^{\rtm{R}}_2
\end{array}
\right).
\end{equation}
Here, each term of the diagonal and non-diagonal blocks is a $2\times2$ matrix, whose implicit form directly follows from the \textit{ab initio} ${\mathbf k}\cdot{\mathbf p}$ expansion up to the third order in \textbf{k}: $E_n=\epsilon_n\rtm{\mathbb{I}}_{2\times2}$, with $\rtm{\mathbb{I}}_{2\times2}$ being the $2\times2$ identity matrix, $H_n=M_{n}k^2\rtm{\mathbb{I}}_{2\times2}$, $k=\sqrt{k_x^2+k_y^2}$, and
\begin{equation}\label{Rashba2}
H^{\rtm{R}}_n=\left(
\begin{array}{cc}
-iW_n (k_+^3-k_-^3)      & i\tilde{\alpha}_{n}k_- \\
-i\tilde{\alpha}_{n}k_+ & iW_n(k_+^3-k_-^3)
\end{array}
\right),
\end{equation}
where $\tilde{\alpha}_{n}=\alpha_n^{(1)}+\alpha_n^{(3)}k^2$ and $k_{\pm}=k_x\pm k_y$. The well-known $2\times2$ Rashba term $H^{\rtm{R}}_n$ is responsible for the out-of-plane and in-plane spin structure typical of hexagonal structures, see, e.g, Ref.~[\onlinecite{Nechaev_PRB_2019}] and references therein. The interaction between the states $n=1$ and 2 is realized through the term
\begin{equation}\label{HintTWOSS}
\widetilde{H}=\left(
\begin{array}{cc}
i\theta k_+^3+i\eta k_-^3      & i\tilde{\zeta}k_- +Dk_+^2 \\
-i\tilde{\zeta}k_+ -Dk_-^2 & i\theta k_-^3+i\eta k_+^3
\end{array}
\right)
\end{equation}
with $\tilde{\zeta}=\zeta^{(1)}+\zeta^{(3)}k^2$.

\begin{table}
\caption{\label{tab:table1} Parameters of the four-band Hamiltonians~(\ref{TWOSSHam}) (based on calculations for 6QL-layer slabs with the lattice parameter $a=7.8187$~a.u.  for Bi$_2$Se$_3$ and $a=8.0880$~a.u. for Bi$_2$Te$_2$Se). All parameters are in Rydberg atomic units except for $\epsilon_1$ and $\epsilon_2$ presented in eV.}
\begin{ruledtabular}
\begin{tabular}{ldd}
                                                            &       \mbox{Bi$_2$Se$_3$}     &  \mbox{Bi$_2$Te$_2$Se}  \\
\hline
$\epsilon_1$                                    &   -0.118                   &   -0.278   \\
$\epsilon_2$                                    &    1.429                    &   1.053  \\
$\alpha_1^{(1)}$                           &    0.174                     &   0.187  \\
$\alpha_2^{(1)}$                          &    -0.265                     &   -0.150  \\
$\alpha_1^{(3)}$                          &     28.30                    &  -29.03     \\
$\alpha_2^{(3)}$                          &    141.80                  &   -22.35     \\
$\theta$                                           &   8.33                       &   22.39  \\
$\eta$                                              &    -1.02                        &  -3.00 \\
$\zeta^{(1)}$                                 &    -0.048                     &    -0.078 \\
$\zeta^{(3)}$                                 &    -52.83                     &    4.51   \\
$D$                                                   &   -2.64                       &   -3.40 \\
$M_1$                                               &   7.97                        &    15.67 \\
$M_2$                                               &  -2.56                        &    -1.76 \\
$M_0$                                               &  -0.37                        &    0.61 \\
$W_1$                                               &  -4.71                        &  -17.32 \\
$W_2$                                               &   5.53                       &    11.88  \\
$s^{\spr}_1$                                    &    0.70                       &   0.63  \\
$s^{\spr}_2$                                    &    0.42                       &   0.39  \\
$ s^{z}_1$                                        &     0.40                       &   0.26 \\
$ s^{z}_2$                                        &   -0.16                       &   -0.21 \\
$\tilde{s}^{\spr}$                          &  -0.21                        &  -0.10  \\
$\tilde{s}^{z}$                                &  -0.41                       &   -0.20 \\
\end{tabular}
\end{ruledtabular}
\end{table}

In the new basis, the spin matrix that yields the spin structure of the states under study is defined as 
\begin{equation}\label{SpinTWOSS}
\rtm{\mathbf{S}}^{4\times4}_{\rtm{\mathbf{kp}}}=\left(
\begin{array}{cc}
\rtm{\mathbf{S}}_1 & \widetilde{\rtm{\mathbf{S}}}  \\
\widetilde{\rtm{\mathbf{S}}} & \rtm{\mathbf{S}}_2
\end{array}
\right)
\end{equation}
with $\rtm{\mathbf{S}}_n=(s^{\spr}_n\bm{\sigma}_{\spr}, s^{z}_n\sigma_z)$ and $\widetilde{\rtm{\mathbf{S}}}=(\tilde{ s}^{\spr}\bm{\sigma}_{\spr}, \tilde{s}^{z}\sigma_z)$, where  $\bm{\sigma}_{\spr} = (\sigma_x, \sigma_y)$ and $\sigma_x$, $\sigma_y$, and $\sigma_z$ are the Pauli matrices. The elements of the spin matrix $\left[\rtm{\mathbf{S}}^{4\times4}_{\rtm{\mathbf{kp}}} \right]^{n\mu}_{l\nu} = \langle\Phi_{n\mu}|\bm{\sigma}|\Phi_{l\nu}\rangle$ enter into the expression for the spin expectation value
\begin{equation}\label{modelRealSpin}
\langle \mathbf{S}_{\mathbf{k}\lambda}\rangle = \frac{1}{2} \langle \widetilde{\Phi}^{\lambda}_{\mathbf{k}}|\bm{\sigma}|\widetilde{\Phi}^{\lambda}_{\mathbf{k}}\rangle
= \frac{1}{2}\sum\limits_{n\mu l\nu} C_{{\mathbf{k}}n\mu}^{\lambda\ast}C_{{\mathbf{k}}l\nu}^{\lambda}\left[\rtm{\mathbf{S}}^{4\times4}_{\rtm{\mathbf{kp}}} \right]^{n\mu}_{l\nu}
\end{equation}
in the state $|\widetilde{\Phi}^{\lambda}_{\mathbf{k}}\rangle =\sum\limits_{n\mu}C_{\mathbf{k}n\mu}^{\lambda} |\Phi_{n\mu}\rangle$ of the reduced Hilbert space of the Hamiltonian $H^{4\times4}_{\rtm{\mathbf{kp}}}$. The four-dimensional vectors $\mathbf{C}^{\lambda}_{\mathbf{k}}$ diagonalize this Hamiltonian  $H^{4\times4}_{\rtm{\mathbf{kp}}} \mathbf{C}^{\lambda}_{\mathbf{k}} = E^{\lambda}_{\mathbf{k}} \mathbf{C}^{\lambda}_{\mathbf{k}}$. The parameters in Eqs.~(\ref{TWOSSHam}) and (\ref{SpinTWOSS}) are listed in Table~\ref{tab:table1}. The bands obtained with these parameters are shown in Fig.~\ref{fig1} by green lines.

Next, we \textit{analytically} transform the Hamiltonian~(\ref{TWOSSHam}) by means of the L\"{o}wdin partitioning, retaining terms up to seventh-order in $\mathbf{k}$ for the block $E_1+H_1+H^{\rtm{R}}_1$ of this Hamiltonian. As a result, we arrive at the $2\times2$ Hamiltonian that describes the low-energy Dirac surface state:
\begin{widetext}
\begin{equation}\label{UPTO7}
H_{\rtm{\mathbf{kp}}}^{2\times2}=\left(
\begin{array}{cc}
\epsilon_1+\mathcal{M}k^2-i\mathcal{W}(k_+^3-k_-^3)+\mathcal{N}(k_+^6+k_-^6)  & i\tilde{\alpha}k_--i\tilde{\gamma}k_+^5+i\xi k_-^7 \\
-i\tilde{\alpha}k_++i\tilde{\gamma}k_-^5-i\xi k_+^7  & \epsilon_1+\mathcal{M}k^2+i\mathcal{W}(k_+^3-k_-^3)+\mathcal{N}(k_+^6+k_-^6)
\end{array}
\right),
\end{equation}
\end{widetext}
where $\mathcal{M}=\sum_{m=0}^{2}M^{(2m)}k^{2m}$, $\mathcal{W}=\sum_{m=0}^{2}W^{(2m)}k^{2m}$, $\tilde{\alpha}=\alpha_1^{(1)}+\sum_{m=1}^{3}\alpha^{(2m+1)}k^{2m}$, and $\tilde{\gamma}=\gamma^{(5)}+\gamma^{(7)}k^2$, see Appendix~\ref{A_params}. All the parameters are listed in Table~\ref{tab:table2}. With these parameters, the diagonalization of the Hamiltonian~(\ref{UPTO7}) yields the bands shown by dark green lines in Fig.~\ref{fig1}.

Up to fifth order in \textbf{k}, the Hamiltonian~(\ref{UPTO7}) of our \textit{ab initio} ${\mathbf k}\cdot{\mathbf p}$ theory is in accord with the form of  the two-band Hamiltonian constructed in Ref.~[\onlinecite{Basak_PRB_2011}] for Bi$_2$Te$_3$ considering the $C_{3v}$ crystal symmetry and time-reversal symmetry. In Ref.~[\onlinecite{Hopfner_PRL_2012}], the Hamiltonian of Ref.~[\onlinecite{Basak_PRB_2011}] was modified by adding a sixth-order term $k_+^6+k_-^6$ in order to reproduce the hexagonal warping of the Au/Ge(111) surface state not related to the spin-orbit effect. Obviously, this term is naturally present in our theory. Note that in Refs.~[\onlinecite{Basak_PRB_2011}] and [\onlinecite{Hopfner_PRL_2012}] the values of  the parameters were found by fitting the model Hamiltonian to \textit{ab initio} results, and, for example, in the case of the surface state of Bi$_2$Te$_3$~[\onlinecite{Basak_PRB_2011}] the values for the lower-order terms differ strongly from those obtained in Refs.~[\onlinecite{Liu_PRB_2010}] and [\onlinecite{Nechaev_PRBR_2016}], which are currently commonly accepted. In contrast to a fitting method, in our $\mathbf{k}\cdot\mathbf{p}$ theory the shape and the value of a given order term are independent on whether or not we include higher-order terms (and it does not affect the lower-order terms), since our $\mathbf{k}\cdot\mathbf{p}$ expansion uniquely follows from the basis set---the eigenfunctions of the original {\it ab initio} Hamiltonian.

\begin{table}
\caption{\label{tab:table2} Parameters of the two-band Hamiltonians~(\ref{UPTO7}) in Rydberg atomic units. The parameters $\epsilon_1$, $\alpha_1^{(1)}$, $s^{\spr}_1$, $ s^{z}_1$, and the lattice parameters $a$ are listed in Table~\ref{tab:table1}. }
\begin{ruledtabular}
\begin{tabular}{ldd}
                                                            &       \mbox{Bi$_2$Se$_3$}     &  \mbox{Bi$_2$Te$_2$Se}  \\
\hline
$\alpha^{(3)}$                               & 27.91                      &    -28.27    \\
$\alpha^{(5)}$                               &  -575.48                    &  115.57     \\
$\alpha^{(7)}$                               &  -92731.72               &   63344.71  \\
$\gamma^{(5)}$                            & 529.40                   &   1735.90    \\
$\gamma^{(7)}$                            &  -39773.06              &   319791.93  \\
$\xi$                                                  &  51.10                    &    589.48  \\
$\mathcal{N}$                               &  -157.85                   &     -789.83\\
$\mathcal{M}^{(0)}$                   &   7.95                         &     15.61 \\
$\mathcal{M}^{(2)}$                   &   -110.14                  &    -122.40 \\
$\mathcal{M}^{(4)}$                   & -36925.53                &    -27778.63 \\
$\mathcal{W}^{(0)}$                   &  -5.82                         &     -20.03 \\
$\mathcal{W}^{(2)}$                   &  -1389.00                  &     -162.47 \\
$\mathcal{W}^{(4)}$                   &   -111467.25            &     65344.75 \\
\end{tabular}
\end{ruledtabular}
\end{table}

The spin-resolved CECs and the non-orthogonality by our ${\mathbf k}\cdot{\mathbf p}$ model are shown in Fig.~\ref{fig2}. As seen in the figure, the effective model underestimates the non-orthogonality for the lower-energy Dirac surface states. A better agreement with the respective \textit{ab initio} results is achieved by increasing the magnitude of $\gamma^{(5)}$  by a factor of 4.5 ($\gamma^{(5)}=2382.3$~a.u.) and 3.8 ($\gamma^{(5)}=6596.4$~a.u.) for Bi$_2$Se$_3$ and Bi$_2$Te$_2$Se, respectively, see the orange areas and curves in Fig.~\ref{fig2} (the respective energy bands are shown by orange lines in Fig.~\ref{fig1}). Note that by manually correcting this parameter we reproduce more accurately not only the non-orthogonality, but also the hexagonal warping of the contours.

\section{Effective fields and multiple winding}

\begin{figure}[tbp]
\centering
\includegraphics[width=\columnwidth]{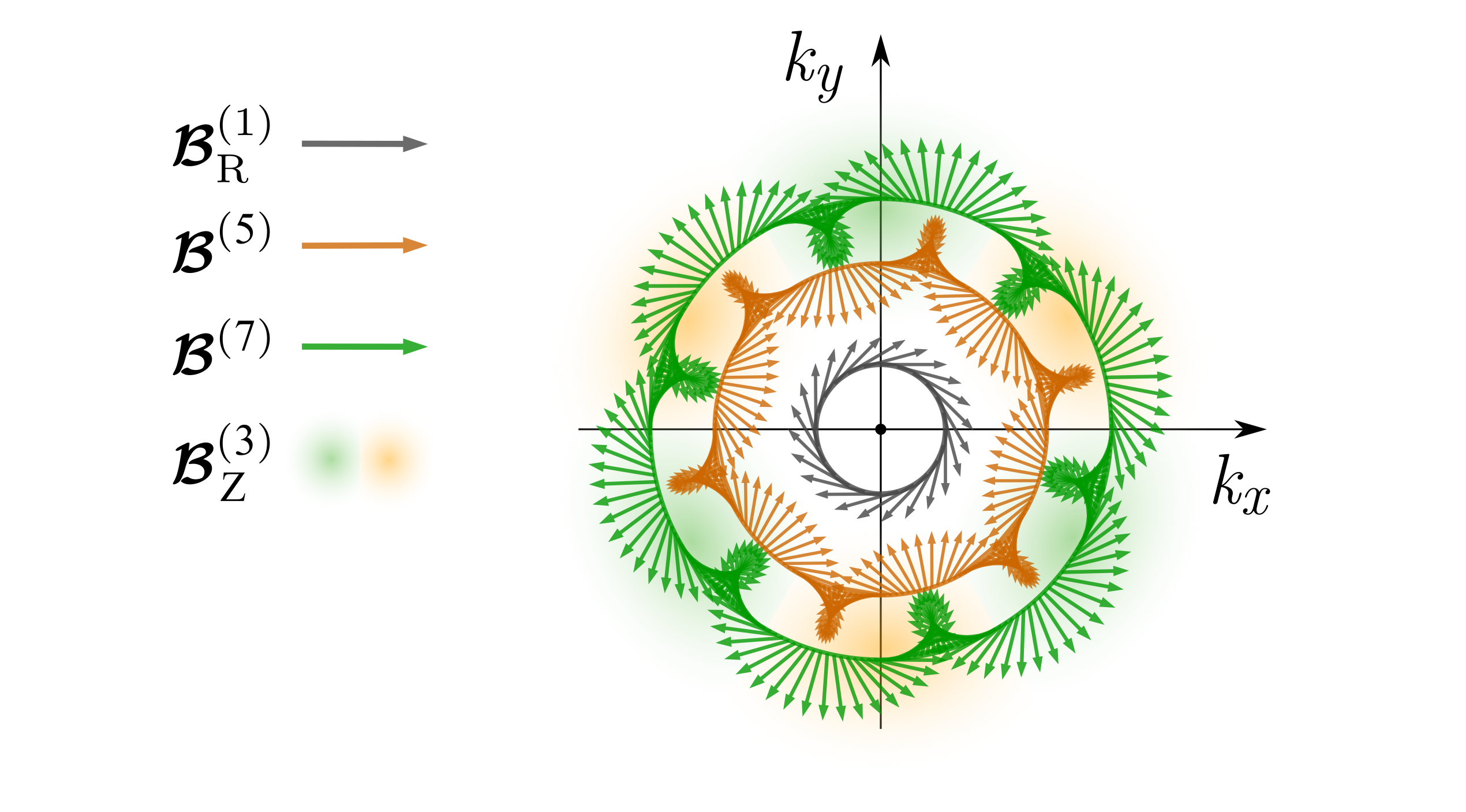}
\caption{Effective spin-orbit magnetic fields in Eq.~(\ref{TwoBandEMF}) as a function of polar angle $\varphi_{{\mathbf k}}$.  The in-plane fields ${\bm{\mathcal{B}}}_{\rtm{R}}^{(1)}$,  ${\bm{\mathcal{B}}}^{(5)}$, and ${\bm{\mathcal{B}}}^{(7)}$ are represented by arrows showing the direction of the field vectors at a given ${\mathbf k}$. The out-of-plane field ${\bm{\mathcal{B}}}_{\rtm{Z}}^{(3)}$ is illustrated by the blurred sixty-degree sectors of green and orange color for the negative and positive sign of its $z$ projection, respectively.}
\label{fig3}
\end{figure}

We focus now on the terms of the $2\times2$ Hamiltonian that cause the non-orthogonality in the in-plane-spin structure. We rewrite the Hamiltonian~(\ref{UPTO7}) in terms of the Pauli matrices:
\begin{equation}\label{TwoBandHam}
  H^{2\times2}_{\rtm{\mathbf{kp}}} = \mathcal{E}(\mathbf{k})\sigma_0 + {\bm{\mathcal{B}}}\cdot{\bm\sigma},
\end{equation}
where $\mathcal{E}(\mathbf{k}) = \epsilon_1 + \mathcal{M}k^2+2\mathcal{N}k^6\cos6\varphi_{{\mathbf k}}$ represents the dispersion of the doubly degenerate bands with the hexagonal warping of their CECs. The SOI-induced splitting of the bands
\begin{equation}\label{HamEig}
E^{\pm}(\mathbf{k}) = \mathcal{E}(\mathbf{k}) \pm |\bm{\mathcal{B}}|
\end{equation}
is due to the Zeeman-like term with the effective (spin-orbit) magnetic field
\begin{equation}\label{TwoBandEMF}
{\bm{\mathcal{B}}} = \widetilde{\alpha}{\bm{\mathcal{B}}}_{\rtm{R}}^{(1)} + 2\mathcal{W}{\bm{\mathcal{B}}}_{\rtm{Z}}^{(3)} + \widetilde{\gamma}{\bm{\mathcal{B}}}^{(5)} +\xi{\bm{\mathcal{B}}}^{(7)}.
\end{equation}
This field consists of the classical (linear) Rashba magnetic field ${\bm{\mathcal{B}}}_{\rtm{R}}^{(1)} =  k(\sin\varphi_{{\mathbf k}},-\cos\varphi_{{\mathbf k}},0)$, the cubic field ${\bm{\mathcal{B}}}_{\rtm{Z}}^{(3)} =  k^3(0,0,\sin3\varphi_{{\mathbf k}})$ responsible for the well-known three-fold  symmetric pattern of the spin $z$ component and contributing to the hexagonal warping of the CECs~[\onlinecite{Fu_PRL_2009}], and two higher-order six-fold symmetric fields ${\bm{\mathcal{B}}}^{(5)} =  k^5(\sin5\varphi_{{\mathbf k}},\cos5\varphi_{{\mathbf k}},0)$ and ${\bm{\mathcal{B}}}^{(7)}  =  k^7(\sin7\varphi_{{\mathbf k}}, - \cos7\varphi_{{\mathbf k}},0)$, see Fig.~\ref{fig3}. Note that since the spin matrix $(s^{\spr}_1\bm{\sigma}_{\spr}, s^{z}_1\sigma_z)$ of our two-band ${\mathbf k}\cdot{\mathbf p}$ model [the upper-left $2\times2$ block of the spin matrix~(\ref{SpinTWOSS})] differs from the matrix $(\bm{\sigma}_{\spr}, \sigma_z)$ of a model built on a scalar-relativistic basis only by the non-unity coefficients $s^{\spr}_1$ and $s^{z}_1$, one is tempted to treat the Pauli matrices in Eq.~(\ref{TwoBandHam}) as if they were spin matrices. Then, the spin expectation value is $\mathbf{S}^{\pm}({\mathbf{k}})=\pm \frac{1}{2}\bm{\mathcal{B}}/ |\bm{\mathcal{B}}|$. However, irrespective of the interpretation of $\bm{\sigma}$ in Eq.~(\ref{TwoBandHam}), in our model the non-orthogonality is characterized by the deviation angle $\delta^{\pm}$ found from the dot product
\begin{equation}\label{delta_nonorth}
\sin\delta^{\pm} = \frac{\mathbf{S}^{\pm}_{\spr}({\mathbf{k}})\cdot \mathbf{k}} {|\mathbf{S}^{\pm}_{\spr}({\mathbf{k}})|k}  = \pm \frac{k^5}{ |\bm{\mathcal{B}}_{\spr}|}(\widetilde{\gamma}+\xi k^2)\sin6\varphi_{\mathbf{k}},
\end{equation}
where the parallel (in-plane) component of the effective field~(\ref{TwoBandEMF}) is
\begin{eqnarray}\label{EMF_parallel}
|\bm{\mathcal{B}}_{\spr}|^2&=&(\widetilde{\alpha}^2+\widetilde{\gamma}^2k^8+ \xi^2k^{12})k^2 \nonumber \\
&-& 2\widetilde{\alpha} k^6(\widetilde{\gamma}-\xi k^2)\cos6\varphi_{\mathbf{k}} \\
&-&2\widetilde{\gamma}\xi k^{12} \cos12\varphi_{\mathbf{k}}. \nonumber
\end{eqnarray}
We see that the angle $\delta^{\pm}$ has a non-trivial dependence on the polar angle with a $\pi/3$ periodicity due to the presence of the fields ${\bm{\mathcal{B}}}^{(5)}$ and ${\bm{\mathcal{B}}}^{(7)}$. Acting separately, these fields may cause a quintuple or a septuple winding of the in-plane spin, respectively, in contrast to the Rashba field ${\bm{\mathcal{B}}}_{\rtm{R}}^{(1)}$ yielding a single winding, Fig.~\ref{fig3}.

At a given $\mathbf{k}$, the importance of each contribution to the effective magnetic field of Eq.~(\ref{TwoBandEMF}) depends on the respective parameter of the Hamiltonian~(\ref{UPTO7}):  $\widetilde{\alpha}$, $\mathcal{W}$, $\widetilde{\gamma}$ or $\xi$. According to their values in Table~\ref{tab:table2}, the effect of the field ${\bm{\mathcal{B}}}^{(7)}$ is expected to be negligible, because $\xi$ is much smaller than $\alpha^{(7)}$ and $\gamma^{(7)}$. At the same time, the contribution of ${\bm{\mathcal{B}}}^{(5)}$ depends on the parameters $\gamma^{(5)}$ and $\gamma^{(7)}$, which are comparable to or even larger than $\alpha^{(5)}$ and $\alpha^{(7)}$, respectively. However, because of the dominant contribution of the linear Rashba field ${\bm{\mathcal{B}}}_{\rtm{R}}^{(1)}$ due to the rather large $\alpha^{(1)}_1$ in $\widetilde{\alpha}$ of Eq.~(\ref{TwoBandEMF}), the superposition of all the in-plane fields produces a single winding of the in-plane spin~[\onlinecite{Note1}].

The in-plane-filed contribution~(\ref{EMF_parallel}) as well as the out-of-plane contribution $|{\bm{\mathcal{B}}}_{Z}|^2 = 2 \mathcal{W}^2 k^6 (1-\cos6\varphi_{\mathbf{k}})$ of the effective field~(\ref{TwoBandEMF}) affects the eigenvalues~(\ref{HamEig}) of the Hamiltonian~(\ref{TwoBandHam}) though the splitting term $\pm|{\bm{\mathcal{B}}}|$. This means that the SOI-induced hexagonal warping of the CECs is due not only to the cubic field ${\bm{\mathcal{B}}}_{Z}$ as, e.g., in Ref.~[\onlinecite{Fu_PRL_2009}], but also to the fields ${\bm{\mathcal{B}}}^{(5)}$ and ${\bm{\mathcal{B}}}^{(7)}$, which contribute to the warping through the scalar products ${\bm{\mathcal{B}}}^{(5)}\cdot{\bm{\mathcal{B}}}_{\rtm{R}}^{(1)}$ and ${\bm{\mathcal{B}}}^{(7)}\cdot{\bm{\mathcal{B}}}_{\rtm{R}}^{(1)}$ [the terms proportional to $\widetilde{\alpha}\widetilde{\gamma}$ and $\widetilde{\alpha}\xi$ in Eq.~(\ref{EMF_parallel}), respectively]. As follows form Tables~\ref{tab:table1} and \ref{tab:table2}, the fields ${\bm{\mathcal{B}}}_{\spr}$ and ${\bm{\mathcal{B}}}_{\rtm{Z}}$ are equally important for the $\cos6\varphi_{\mathbf{k}}$ distortion of the CEC. In addition, the hexagonal warping due to ${\bm{\mathcal{B}}}_{\rtm{Z}}$ gives rise to the $z$ spin component, so if one neglects the contribution of ${\bm{\mathcal{B}}}_{\spr}$ and fits only the cubic field to calculated or measured CECs, one may arrive at a large out-of-plane spin polarization with the spin-momentum locking unaffected by the warping. In contrast, the CEC warping caused by the fields ${\bm{\mathcal{B}}}^{(5)}$ and ${\bm{\mathcal{B}}}^{(7)}$ is accompanied by a change of the locking angle between spin and momentum. This explains why a stronger  warping may imply a larger non-orthogonality.

\section{Conclusions}

To summarize, within a fully \textit{ab initio} ${\mathbf k}\cdot{\mathbf p}$ perturbation approach we have developed a two-band effective ${\mathbf k}\cdot{\mathbf p}$ model for the surface states of the three-dimensional topological insulators Bi$_2$Se$_3$ and Bi$_2$Te$_2$Se. The model includes terms to seventh order in $\mathbf{k}$ and provides a comprehensive description of the surface spin structure characterized by a non-orthogonality between the surface-electron spin and its momentum. In the ${\mathbf k}\cdot{\mathbf p}$ theory, the non-orthogonality that arises naturally in the \textit{ab initio} calculations is included in the effective models through the higher-order terms in \textbf{k}. Our $\mathbf{k}\cdot\mathbf{p}$ expansion builds on the eigenfunctions of the {\it ab initio} Hamiltonian, and, therefore, a term of a given order is unambiguously determined by the \textit{ab initio} spinor wave functions and, in contrast to a fitting method, does not depend on the presence of other terms. We have shown that the $k_{\pm}^5$- and $k_{\pm}^7$-terms represent effective spin-orbit magnetic fields with six-fold symmetric patterns on the two-dimensional momentum plane and, thereby, can lead to a non-orthogonality with the $\pi/3$ periodicity as a function of the polar angle of $\mathbf{k}$. For Bi$_2$Se$_3$ and Bi$_2$Te$_2$Se, we have found that the contribution of the $k_{\pm}^7$-term is rather small, and it is the $k_{\pm}^5$-term that causes the few-degree deviation of the actual spin direction from the classical orthogonality.

Finally, we would like to note that the derived two-band Hamiltonian is fully applicable to classical Rashba systems such as the Au(111) surface state. Here, the non-orthogonality appears to be negligibly small, albeit nonzero. A similar study for the giant Rashba spin-split conduction state of a single BiTeI trilayer reveals a substantial non-orthogonality, rather different for the inner and outer constant energy contour. In fact, the two-band Hamiltonian (\ref{UPTO7}) can be considered typical of hexagonal structures. Thus, the simplified picture that the in-plane spin and momentum are locked perpendicular to each other by spin-orbit interaction might overlook important features inherent in the spin-related phenomena at the surfaces and in 2D structures.

\begin{acknowledgments}
This work was supported by funding from the Department of Education of the Basque Government under Grant No.~IT1164-19 and by the Spanish Ministry of Science, Innovation and Universities (Project No.~FIS2016-76617-P).
\end{acknowledgments}

\appendix

\section{\textit{Ab initio} third-order ${\mathbf k}\cdot{\mathbf p}$ expansion}\label{A_expansion}

The L\"{o}wdin partitioning applied to the original Hilbert space of the LDA Hamiltonian, represents the ${\mathbf k}\cdot{\mathbf p}$ Hamiltonian in the basis of the chosen spinor wave functions (the states in set $A$ numbered below by the indices $n\nu$, $m\mu$, and $m'\mu'$) in terms of the matrix elements of the velocity operator\cite{Nechaev_PRBR_2016, Krasovskii_PRB_2014}
$$
\bm{\pi}=-i\hbar\mathrm{\bm{\nabla}}+\frac{\hbar}{4m_0c^2}\left[\bm{\sigma} \times \mathrm{\bm{\nabla}} V\right]
$$
Here, $\bm{\sigma}$ is the vector of the Pauli matrices that operate on spinors, and $V(\mathrm{\mathbf{r}})$ is the crystal potential. The \textit{ab initio} third-order ${\mathbf k}\cdot{\mathbf p}$ expansion at a TRIM in a centrosymmetric system\cite{Nechaev_PRB_2019, Usachov_arXiv_2020} reads
$$
H_{\mathrm{\mathbf{kp}}}=H^{(0)}+H^{(1)}+H^{(2)}+H^{(3)},
$$
where the zero-order term is just the band energy,
$$
H^{(0)}_{n\nu m\mu}=\epsilon_n\delta_{mn}\delta_{\nu\mu},
$$
and the linear term is
$$
H^{(1)}_{n\nu m\mu}=\frac{\hbar}{m_0}\mathbf{k}\cdot\bm{\pi}_{n\nu m\mu}
$$
with the matrix elements $\bm{\pi}_{n\nu m\mu}=\langle \Psi_{n\nu}|\bm{\pi}|\Psi_{ m\mu}\rangle$. For two Kramers pairs of different parity, $n$th and $m$th, we turn the phases such that $i\pi^{x(z)}_{n\uparrow m\downarrow}$ and/or $i\pi^{x(z)}_{n\uparrow m\uparrow}$ be real.
The  second- and third-order terms are
\begin{eqnarray*}
H^{(2)}_{n\nu m\mu}&=&\frac{\hbar^2k^2}{2m_0}\delta_{mn}\delta_{\nu\mu}+\frac{\hbar^2}{m^2_0} \sum_{\alpha\beta}k_{\alpha}k_{\beta}D_{n\nu m\mu}^{\alpha\beta},\\
H^{(3)}_{n\nu m\mu}&=&\frac{\hbar^3}{m^3_0}\sum_{\alpha\beta\gamma} k_{\alpha}k_{\beta}k_{\gamma}T_{n\nu m\mu}^{\alpha\beta\gamma}
\end{eqnarray*}
with $\alpha,\beta,\gamma=x,y,z$. Here, the coefficients are
\begin{eqnarray*}
D_{n\nu m\mu}^{\alpha\beta}&=&\frac{1}{2}\sum_{ l\eta}\pi^{\alpha}_{n\nu l\eta} \pi^{\beta}_{ l\eta m\mu}\left(\frac{1}{\Delta_{nl}}+\frac{1}{\Delta_{ml}}\right),\\
T_{n\nu m\mu}^{\alpha\beta\gamma}&=&-\frac{1}{2}\sum_{ l\eta, m'\mu'} \left[\frac{\pi^{\alpha}_{n\nu l\eta}\pi^{\beta}_{ l\eta m'\mu'}\pi^{\gamma}_{m'\mu' m\mu}}
{\Delta_{ml}\Delta_{m'l}}\right.\\
&+&\left.\frac{\pi^{\alpha}_{n\nu m'\mu'}\pi^{\beta}_{ m'\mu' l\eta}\pi^{\gamma}_{ l\eta m\mu}}
{\Delta_{nl}\Delta_{m'l}}\right]\\
&+&\frac{1}{2}\sum_{ l\eta, l'\eta'}\pi^{\alpha}_{n\nu l\eta}\pi^{\beta}_{ l\eta l'\eta'}\pi^{\gamma}_{ l'\eta' m\mu}\\
&\times&\left[\frac{1} {\Delta_{nl}\Delta_{nl'}}+\frac{1}{\Delta_{ml}\Delta_{ml'}}\right],
\end{eqnarray*}
where $\Delta_{nl}=\epsilon_n-\epsilon_l$, and the indices $l\eta$ and $l'\eta'$ number the states in set $B$, i.e., run over all the states of the original Hilbert space excluding those forming the ${\mathbf k}\cdot{\mathbf p}$ basis---the subspace $A$.

\section{Parameters of the $2\times2$ Hamiltonian}\label{A_params}

The analytical transformation of the four-band Hamiltonian (\ref{TWOSSHam}) by means of the L\"{o}wdin partitioning leads to the following expressions for the parameters of the two-band Hamiltonian~(\ref{UPTO7}):
\begin{widetext}

\begin{eqnarray*}
\alpha^{(3)}&=&\alpha_1^{(3)}-\frac{2M_0\zeta^{(1)}}{\Delta_{21}} -\frac{\zeta^{(1)2}}{\Delta_{21}^2}[\alpha^{(1)}_1-\alpha^{(1)}_2],\\
\alpha^{(5)}&=&-\frac{2}{\Delta_{21}}[M_0\zeta^{(3)}-\eta D]-\frac{1}{\Delta_{21}^2}\left( [M_0^2+2\zeta^{(1)}\zeta^{(3)}] [\alpha^{(1)}_1-\alpha^{(1)}_2] + \alpha_1^{(1)}D^2+2M_0\zeta^{(1)}[M_1-M_2]+ \zeta^{(1)2}[\alpha_1^{(3)}-\alpha_2^{(3)}]\right),\\
\alpha^{(7)}&=& \frac{1}{\Delta_{21}^2}\left( 2[M_1-M_2][\eta D-M_0\zeta^{(3)}]-2W_2[M_0D+\zeta^{(1)}(\theta-\eta)] \right)\nonumber\\
&+&\frac{1}{\Delta_{21}^2}\left( [\alpha_1^{(3)}-\alpha_2^{(3)}][M_0^2+2\zeta^{(1)}\zeta^{(3)}]- [\alpha_1^{(1)}-\alpha_2^{(1)}]\zeta^{(3)2}-\alpha^{(3)}_1D^2-\alpha_1^{(1)} (\theta^2+\eta^2)+2\alpha_2^{(1)}\theta\eta\right),\nonumber\\
\end{eqnarray*}

\begin{eqnarray*}
W^{(0)}&=&W_1-\zeta^{(1)}D/\Delta_{21},\\
W^{(2)}&=&\frac{1}{\Delta_{21}}\left(M_0(\theta-\eta)-\zeta^{(3)}D\right)-\frac{1}{\Delta_{21}^2}\left(\zeta^{(1)2}[W_1+W_2] +\zeta^{(1)}D[M_1-M_2]+\alpha_2^{(1)}\zeta^{(1)}[\theta-\eta]-\alpha_2^{(1)}M_0D\right),\\
W^{(4)}&=& \frac{1}{\Delta_{21}^2}\left( (\theta - \eta)[M_0(M_1-M_2)-\alpha_2^{(1)}\zeta^{(3)} - \alpha_2^{(3)}\zeta^{(1)}]-\zeta^{(3)}D[M_1-M_2] \right) \nonumber\\
 &-&\frac{1}{\Delta_{21}^2}\left( M_0^2[W_1-W_2]+ [W_1+W_2][D^2+2\zeta^{(1)}\zeta^{(3)}] -\alpha^{(3)}M_0D \right), \nonumber\\
\end{eqnarray*}

\begin{eqnarray*}
M^{(0)}&=&M_1-\zeta^{(1)2}/\Delta_{21},\\
M^{(2)}&=&-\frac{1}{\Delta_{21}}\left(M_0^2+D^2+2\zeta^{(1)}\zeta^{(3)}\right)-\frac{1}{\Delta_{21}^2}\left(\zeta^{(1)2}[M_1-M_2]+ 2\zeta^{(1)}M_0[\alpha^{(1)}_1-\alpha^{(1)}_2]\right),\\
M^{(4)}&=&-\frac{1}{\Delta_{21}}\left(\theta^2+\eta^2+\zeta^{(3)2} \right) -\frac{1}{\Delta_{21}^2}\left( [M_1-M_2][M_0^2+D^2+2\zeta^{(1)}\zeta^{(3)}] \right) \nonumber\\
   &-&\frac{2}{\Delta_{21}^2}\left(\zeta^{(3)}M_0 [\alpha_1^{(1)}-\alpha_2^{(1)}] -D(\alpha_1^{(1)}\eta-\alpha_2^{(1)} \theta)+ \zeta^{(1)}M_0[\alpha_1^{(3)}-\alpha_2^{(3)}] +\zeta^{(1)}D[W_1+W_2]\right),  \nonumber\\
\end{eqnarray*}

\begin{eqnarray*}
\gamma^{(5)}&=&-\frac{2\theta D}{\Delta_{21}}-\frac{D^2\alpha^{(1)}_2}{\Delta_{21}^2},\\
\gamma^{(7)}&=&-\frac{1}{\Delta_{21}^2}\left( 2\theta D[M_1-M_2]+2W_2[M_0D+\theta\zeta^{(1)}] -\alpha_1^{(1)}\theta\eta +\alpha_2^{(1)}\theta^2 + \alpha_2^{(3)}D^2 \right),\\
\end{eqnarray*}

\begin{eqnarray*}
\mathcal{N}&=&-\frac{\theta\eta}{\Delta_{21}}+\frac{D}{\Delta_{21}^2} \left(\zeta^{(1)}[W_1+W_2]+ \alpha_1^{(1)}\theta-\alpha_2^{(1)}\eta \right),\\
\xi&=&\frac{1}{\Delta_{21}^2}\left(\eta^2\alpha_2^{(1)}-2\eta\zeta^{(1)}W_2-\alpha_1^{(1)}\eta\theta \right),\, \Delta_{21}=\epsilon_2-\epsilon_1.
\end{eqnarray*}
\end{widetext}


\begin{thebibliography}{28}%
\makeatletter
\providecommand \@ifxundefined [1]{%
 \@ifx{#1\undefined}
}%
\providecommand \@ifnum [1]{%
 \ifnum #1\expandafter \@firstoftwo
 \else \expandafter \@secondoftwo
 \fi
}%
\providecommand \@ifx [1]{%
 \ifx #1\expandafter \@firstoftwo
 \else \expandafter \@secondoftwo
 \fi
}%
\providecommand \natexlab [1]{#1}%
\providecommand \enquote  [1]{``#1''}%
\providecommand \bibnamefont  [1]{#1}%
\providecommand \bibfnamefont [1]{#1}%
\providecommand \citenamefont [1]{#1}%
\providecommand \href@noop [0]{\@secondoftwo}%
\providecommand \href [0]{\begingroup \@sanitize@url \@href}%
\providecommand \@href[1]{\@@startlink{#1}\@@href}%
\providecommand \@@href[1]{\endgroup#1\@@endlink}%
\providecommand \@sanitize@url [0]{\catcode `\\12\catcode `\$12\catcode
  `\&12\catcode `\#12\catcode `\^12\catcode `\_12\catcode `\%12\relax}%
\providecommand \@@startlink[1]{}%
\providecommand \@@endlink[0]{}%
\providecommand \url  [0]{\begingroup\@sanitize@url \@url }%
\providecommand \@url [1]{\endgroup\@href {#1}{\urlprefix }}%
\providecommand \urlprefix  [0]{URL }%
\providecommand \Eprint [0]{\href }%
\providecommand \doibase [0]{https://doi.org/}%
\providecommand \selectlanguage [0]{\@gobble}%
\providecommand \bibinfo  [0]{\@secondoftwo}%
\providecommand \bibfield  [0]{\@secondoftwo}%
\providecommand \translation [1]{[#1]}%
\providecommand \BibitemOpen [0]{}%
\providecommand \bibitemStop [0]{}%
\providecommand \bibitemNoStop [0]{.\EOS\space}%
\providecommand \EOS [0]{\spacefactor3000\relax}%
\providecommand \BibitemShut  [1]{\csname bibitem#1\endcsname}%
\let\auto@bib@innerbib\@empty
\bibitem [{\citenamefont {Zhang}\ \emph {et~al.}(2009)\citenamefont {Zhang},
  \citenamefont {Liu}, \citenamefont {Qi}, \citenamefont {Dai}, \citenamefont
  {Fang},\ and\ \citenamefont {Zhang}}]{Zhang_NATPHYS_2009}%
  \BibitemOpen
  \bibfield  {author} {\bibinfo {author} {\bibfnamefont {H.}~\bibnamefont
  {Zhang}}, \bibinfo {author} {\bibfnamefont {C.-X.}\ \bibnamefont {Liu}},
  \bibinfo {author} {\bibfnamefont {X.-L.}\ \bibnamefont {Qi}}, \bibinfo
  {author} {\bibfnamefont {X.}~\bibnamefont {Dai}}, \bibinfo {author}
  {\bibfnamefont {Z.}~\bibnamefont {Fang}},\ and\ \bibinfo {author}
  {\bibfnamefont {S.-C.}\ \bibnamefont {Zhang}},\ }\bibfield  {title} {\bibinfo
  {title} {{Topological insulators in Bi$_2$Se$_3$, Bi$_2$Te$_3$ and
  Sb$_2$Te$_3$ with a single Dirac cone on the surface}},\ }\href
  {https://doi.org/10.1038/nphys1270} {\bibfield  {journal} {\bibinfo
  {journal} {Nature Physics}\ }\textbf {\bibinfo {volume} {5}},\ \bibinfo
  {pages} {438} (\bibinfo {year} {2009})}\BibitemShut {NoStop}%
\bibitem [{\citenamefont {Liu}\ \emph {et~al.}(2010)\citenamefont {Liu},
  \citenamefont {Qi}, \citenamefont {Zhang}, \citenamefont {Dai}, \citenamefont
  {Fang},\ and\ \citenamefont {Zhang}}]{Liu_PRB_2010}%
  \BibitemOpen
  \bibfield  {author} {\bibinfo {author} {\bibfnamefont {C.-X.}\ \bibnamefont
  {Liu}}, \bibinfo {author} {\bibfnamefont {X.-L.}\ \bibnamefont {Qi}},
  \bibinfo {author} {\bibfnamefont {H.}~\bibnamefont {Zhang}}, \bibinfo
  {author} {\bibfnamefont {X.}~\bibnamefont {Dai}}, \bibinfo {author}
  {\bibfnamefont {Z.}~\bibnamefont {Fang}},\ and\ \bibinfo {author}
  {\bibfnamefont {S.-C.}\ \bibnamefont {Zhang}},\ }\bibfield  {title} {\bibinfo
  {title} {{Model Hamiltonian for topological insulators}},\ }\href
  {https://doi.org/10.1103/PhysRevB.82.045122} {\bibfield  {journal} {\bibinfo
  {journal} {Phys. Rev. B}\ }\textbf {\bibinfo {volume} {82}},\ \bibinfo
  {pages} {045122} (\bibinfo {year} {2010})}\BibitemShut {NoStop}%
\bibitem [{\citenamefont {LaShell}\ \emph {et~al.}(1996)\citenamefont
  {LaShell}, \citenamefont {McDougall},\ and\ \citenamefont
  {Jensen}}]{LaShell_PRL_1996}%
  \BibitemOpen
  \bibfield  {author} {\bibinfo {author} {\bibfnamefont {S.}~\bibnamefont
  {LaShell}}, \bibinfo {author} {\bibfnamefont {B.~A.}\ \bibnamefont
  {McDougall}},\ and\ \bibinfo {author} {\bibfnamefont {E.}~\bibnamefont
  {Jensen}},\ }\bibfield  {title} {\bibinfo {title} {{Spin Splitting of an
  Au(111) Surface State Band Observed with Angle Resolved Photoelectron
  Spectroscopy}},\ }\href {https://doi.org/10.1103/PhysRevLett.77.3419}
  {\bibfield  {journal} {\bibinfo  {journal} {Phys. Rev. Lett.}\ }\textbf
  {\bibinfo {volume} {77}},\ \bibinfo {pages} {3419} (\bibinfo {year}
  {1996})}\BibitemShut {NoStop}%
\bibitem [{\citenamefont {Wang}\ \emph {et~al.}(2011)\citenamefont {Wang},
  \citenamefont {Hsieh}, \citenamefont {Pilon}, \citenamefont {Fu},
  \citenamefont {Gardner}, \citenamefont {Lee},\ and\ \citenamefont
  {Gedik}}]{Wang_PRL_2011}%
  \BibitemOpen
  \bibfield  {author} {\bibinfo {author} {\bibfnamefont {Y.~H.}\ \bibnamefont
  {Wang}}, \bibinfo {author} {\bibfnamefont {D.}~\bibnamefont {Hsieh}},
  \bibinfo {author} {\bibfnamefont {D.}~\bibnamefont {Pilon}}, \bibinfo
  {author} {\bibfnamefont {L.}~\bibnamefont {Fu}}, \bibinfo {author}
  {\bibfnamefont {D.~R.}\ \bibnamefont {Gardner}}, \bibinfo {author}
  {\bibfnamefont {Y.~S.}\ \bibnamefont {Lee}},\ and\ \bibinfo {author}
  {\bibfnamefont {N.}~\bibnamefont {Gedik}},\ }\bibfield  {title} {\bibinfo
  {title} {{Observation of a Warped Helical Spin Texture in
  ${\mathrm{Bi}}_{2}{\mathrm{Se}}_{3}$ from Circular Dichroism Angle-Resolved
  Photoemission Spectroscopy}},\ }\href
  {https://doi.org/10.1103/PhysRevLett.107.207602} {\bibfield  {journal}
  {\bibinfo  {journal} {Phys. Rev. Lett.}\ }\textbf {\bibinfo {volume} {107}},\
  \bibinfo {pages} {207602} (\bibinfo {year} {2011})}\BibitemShut {NoStop}%
\bibitem [{\citenamefont {Basak}\ \emph {et~al.}(2011)\citenamefont {Basak},
  \citenamefont {Lin}, \citenamefont {Wray}, \citenamefont {Xu}, \citenamefont
  {Fu}, \citenamefont {Hasan},\ and\ \citenamefont {Bansil}}]{Basak_PRB_2011}%
  \BibitemOpen
  \bibfield  {author} {\bibinfo {author} {\bibfnamefont {S.}~\bibnamefont
  {Basak}}, \bibinfo {author} {\bibfnamefont {H.}~\bibnamefont {Lin}}, \bibinfo
  {author} {\bibfnamefont {L.~A.}\ \bibnamefont {Wray}}, \bibinfo {author}
  {\bibfnamefont {S.-Y.}\ \bibnamefont {Xu}}, \bibinfo {author} {\bibfnamefont
  {L.}~\bibnamefont {Fu}}, \bibinfo {author} {\bibfnamefont {M.~Z.}\
  \bibnamefont {Hasan}},\ and\ \bibinfo {author} {\bibfnamefont
  {A.}~\bibnamefont {Bansil}},\ }\bibfield  {title} {\bibinfo {title} {{Spin
  texture on the warped Dirac-cone surface states in topological insulators}},\
  }\href {https://doi.org/10.1103/PhysRevB.84.121401} {\bibfield  {journal}
  {\bibinfo  {journal} {Phys. Rev. B}\ }\textbf {\bibinfo {volume} {84}},\
  \bibinfo {pages} {121401} (\bibinfo {year} {2011})}\BibitemShut {NoStop}%
\bibitem [{\citenamefont {H\"opfner}\ \emph {et~al.}(2012)\citenamefont
  {H\"opfner}, \citenamefont {Sch\"afer}, \citenamefont {Fleszar},
  \citenamefont {Dil}, \citenamefont {Slomski}, \citenamefont {Meier},
  \citenamefont {Loho}, \citenamefont {Blumenstein}, \citenamefont {Patthey},
  \citenamefont {Hanke},\ and\ \citenamefont {Claessen}}]{Hopfner_PRL_2012}%
  \BibitemOpen
  \bibfield  {author} {\bibinfo {author} {\bibfnamefont {P.}~\bibnamefont
  {H\"opfner}}, \bibinfo {author} {\bibfnamefont {J.}~\bibnamefont
  {Sch\"afer}}, \bibinfo {author} {\bibfnamefont {A.}~\bibnamefont {Fleszar}},
  \bibinfo {author} {\bibfnamefont {J.~H.}\ \bibnamefont {Dil}}, \bibinfo
  {author} {\bibfnamefont {B.}~\bibnamefont {Slomski}}, \bibinfo {author}
  {\bibfnamefont {F.}~\bibnamefont {Meier}}, \bibinfo {author} {\bibfnamefont
  {C.}~\bibnamefont {Loho}}, \bibinfo {author} {\bibfnamefont {C.}~\bibnamefont
  {Blumenstein}}, \bibinfo {author} {\bibfnamefont {L.}~\bibnamefont
  {Patthey}}, \bibinfo {author} {\bibfnamefont {W.}~\bibnamefont {Hanke}},\
  and\ \bibinfo {author} {\bibfnamefont {R.}~\bibnamefont {Claessen}},\
  }\bibfield  {title} {\bibinfo {title} {{Three-Dimensional Spin Rotations at
  the Fermi Surface of a Strongly Spin-Orbit Coupled Surface System}},\ }\href
  {https://doi.org/10.1103/PhysRevLett.108.186801} {\bibfield  {journal}
  {\bibinfo  {journal} {Phys. Rev. Lett.}\ }\textbf {\bibinfo {volume} {108}},\
  \bibinfo {pages} {186801} (\bibinfo {year} {2012})}\BibitemShut {NoStop}%
\bibitem [{\citenamefont {Nechaev}\ and\ \citenamefont
  {Krasovskii}(2016)}]{Nechaev_PRBR_2016}%
  \BibitemOpen
  \bibfield  {author} {\bibinfo {author} {\bibfnamefont {I.~A.}\ \bibnamefont
  {Nechaev}}\ and\ \bibinfo {author} {\bibfnamefont {E.~E.}\ \bibnamefont
  {Krasovskii}},\ }\bibfield  {title} {\bibinfo {title} {{Relativistic
  $\mathrm{k}\ifmmode\cdot\else\textperiodcentered\fi{}\mathrm{p}$ Hamiltonians
  for centrosymmetric topological insulators from \textit{ab initio} wave
  functions}},\ }\href {https://doi.org/10.1103/PhysRevB.94.201410} {\bibfield
  {journal} {\bibinfo  {journal} {Phys. Rev. B}\ }\textbf {\bibinfo {volume}
  {94}},\ \bibinfo {pages} {201410(R)} (\bibinfo {year} {2016})}\BibitemShut
  {NoStop}%
\bibitem [{\citenamefont {Nechaev}\ and\ \citenamefont
  {Krasovskii}(2018)}]{Nechaev_PRB_2018}%
  \BibitemOpen
  \bibfield  {author} {\bibinfo {author} {\bibfnamefont {I.~A.}\ \bibnamefont
  {Nechaev}}\ and\ \bibinfo {author} {\bibfnamefont {E.~E.}\ \bibnamefont
  {Krasovskii}},\ }\bibfield  {title} {\bibinfo {title} {{Relativistic
  splitting of surface states at Si-terminated surfaces of the layered
  intermetallic compounds $R{T}_{2}{\mathrm{Si}}_{2}$ ($R$=rare earth; $T$=Ir,
  Rh)}},\ }\href {https://doi.org/10.1103/PhysRevB.98.245415} {\bibfield
  {journal} {\bibinfo  {journal} {Phys. Rev. B}\ }\textbf {\bibinfo {volume}
  {98}},\ \bibinfo {pages} {245415} (\bibinfo {year} {2018})}\BibitemShut
  {NoStop}%
\bibitem [{\citenamefont {Nechaev}\ and\ \citenamefont
  {Krasovskii}(2019)}]{Nechaev_PRB_2019}%
  \BibitemOpen
  \bibfield  {author} {\bibinfo {author} {\bibfnamefont {I.~A.}\ \bibnamefont
  {Nechaev}}\ and\ \bibinfo {author} {\bibfnamefont {E.~E.}\ \bibnamefont
  {Krasovskii}},\ }\bibfield  {title} {\bibinfo {title} {{Spin polarization by
  first-principles relativistic $\mathrm{k}\cdot\mathrm{p}$ theory: Application
  to the surface alloys ${\mathrm{PbAg}}_{2}$ and ${\mathrm{BiAg}}_{2}$}},\
  }\href {https://doi.org/10.1103/PhysRevB.100.115432} {\bibfield  {journal}
  {\bibinfo  {journal} {Phys. Rev. B}\ }\textbf {\bibinfo {volume} {100}},\
  \bibinfo {pages} {115432} (\bibinfo {year} {2019})}\BibitemShut {NoStop}%
\bibitem [{\citenamefont {Nechaev}\ \emph {et~al.}(2017)\citenamefont
  {Nechaev}, \citenamefont {Eremeev}, \citenamefont {Krasovskii}, \citenamefont
  {Echenique},\ and\ \citenamefont {Chulkov}}]{Nechaev_SciRep_2017}%
  \BibitemOpen
  \bibfield  {author} {\bibinfo {author} {\bibfnamefont {I.~A.}\ \bibnamefont
  {Nechaev}}, \bibinfo {author} {\bibfnamefont {S.~V.}\ \bibnamefont
  {Eremeev}}, \bibinfo {author} {\bibfnamefont {E.~E.}\ \bibnamefont
  {Krasovskii}}, \bibinfo {author} {\bibfnamefont {P.~M.}\ \bibnamefont
  {Echenique}},\ and\ \bibinfo {author} {\bibfnamefont {E.~V.}\ \bibnamefont
  {Chulkov}},\ }\bibfield  {title} {\bibinfo {title} {Quantum spin {Hall}
  insulators in centrosymmetric thin films composed from topologically trivial
  {BiTeI} trilayers},\ }\href {https://doi.org/10.1038/srep43666} {\bibfield
  {journal} {\bibinfo  {journal} {Scientific Reports}\ }\textbf {\bibinfo
  {volume} {7}},\ \bibinfo {pages} {43666} (\bibinfo {year}
  {2017})}\BibitemShut {NoStop}%
\bibitem [{\citenamefont {Schulz}\ \emph {et~al.}(2019)\citenamefont {Schulz},
  \citenamefont {Nechaev}, \citenamefont {G\"{u}ttler}, \citenamefont
  {Poelchen}, \citenamefont {Generalov}, \citenamefont {Danzenb\"{a}cher},
  \citenamefont {Chikina}, \citenamefont {Seiro}, \citenamefont {Kliemt},
  \citenamefont {Vyazovskaya}, \citenamefont {Kim}, \citenamefont {Dudin},
  \citenamefont {Chulkov}, \citenamefont {Laubschat}, \citenamefont
  {Krasovskii}, \citenamefont {Geibel}, \citenamefont {Krellner}, \citenamefont
  {Kummer},\ and\ \citenamefont {Vyalikh}}]{Susanne2019}%
  \BibitemOpen
  \bibfield  {author} {\bibinfo {author} {\bibfnamefont {S.}~\bibnamefont
  {Schulz}}, \bibinfo {author} {\bibfnamefont {I.~A.}\ \bibnamefont {Nechaev}},
  \bibinfo {author} {\bibfnamefont {M.}~\bibnamefont {G\"{u}ttler}}, \bibinfo
  {author} {\bibfnamefont {G.}~\bibnamefont {Poelchen}}, \bibinfo {author}
  {\bibfnamefont {A.}~\bibnamefont {Generalov}}, \bibinfo {author}
  {\bibfnamefont {S.}~\bibnamefont {Danzenb\"{a}cher}}, \bibinfo {author}
  {\bibfnamefont {A.}~\bibnamefont {Chikina}}, \bibinfo {author} {\bibfnamefont
  {S.}~\bibnamefont {Seiro}}, \bibinfo {author} {\bibfnamefont
  {K.}~\bibnamefont {Kliemt}}, \bibinfo {author} {\bibfnamefont {A.~Y.}\
  \bibnamefont {Vyazovskaya}}, \bibinfo {author} {\bibfnamefont {T.~K.}\
  \bibnamefont {Kim}}, \bibinfo {author} {\bibfnamefont {P.}~\bibnamefont
  {Dudin}}, \bibinfo {author} {\bibfnamefont {E.~V.}\ \bibnamefont {Chulkov}},
  \bibinfo {author} {\bibfnamefont {C.}~\bibnamefont {Laubschat}}, \bibinfo
  {author} {\bibfnamefont {E.~E.}\ \bibnamefont {Krasovskii}}, \bibinfo
  {author} {\bibfnamefont {C.}~\bibnamefont {Geibel}}, \bibinfo {author}
  {\bibfnamefont {C.}~\bibnamefont {Krellner}}, \bibinfo {author}
  {\bibfnamefont {K.}~\bibnamefont {Kummer}},\ and\ \bibinfo {author}
  {\bibfnamefont {D.~V.}\ \bibnamefont {Vyalikh}},\ }\bibfield  {title}
  {\bibinfo {title} {{Emerging 2D-ferromagnetism and strong spin-orbit coupling
  at the surface of valence-fluctuating EuIr$_2$Si$_2$}},\ }\href
  {https://doi.org/10.1038/s41535-019-0166-z} {\bibfield  {journal} {\bibinfo
  {journal} {npj Quantum Mater.}\ }\textbf {\bibinfo {volume} {4}},\ \bibinfo
  {pages} {26} (\bibinfo {year} {2019})}\BibitemShut {NoStop}%
\bibitem [{\citenamefont {Usachov}\ \emph {et~al.}(2020)\citenamefont
  {Usachov}, \citenamefont {Nechaev}, \citenamefont {Poelchen}, \citenamefont
  {G\"{u}ttler}, \citenamefont {Krasovskii}, \citenamefont {Schulz},
  \citenamefont {Generalov}, \citenamefont {Kliemt}, \citenamefont {Kraiker},
  \citenamefont {Krellner}, \citenamefont {Kummer}, \citenamefont
  {Danzenb\"{a}cher}, \citenamefont {Laubschat}, \citenamefont {Weber},
  \citenamefont {Chulkov}, \citenamefont {Santander-Syro}, \citenamefont
  {Imai}, \citenamefont {Miyamoto}, \citenamefont {Okuda},\ and\ \citenamefont
  {Vyalikh}}]{Usachov_arXiv_2020}%
  \BibitemOpen
  \bibfield  {author} {\bibinfo {author} {\bibfnamefont {D.~Y.}\ \bibnamefont
  {Usachov}}, \bibinfo {author} {\bibfnamefont {I.~A.}\ \bibnamefont
  {Nechaev}}, \bibinfo {author} {\bibfnamefont {G.}~\bibnamefont {Poelchen}},
  \bibinfo {author} {\bibfnamefont {M.}~\bibnamefont {G\"{u}ttler}}, \bibinfo
  {author} {\bibfnamefont {E.~E.}\ \bibnamefont {Krasovskii}}, \bibinfo
  {author} {\bibfnamefont {S.}~\bibnamefont {Schulz}}, \bibinfo {author}
  {\bibfnamefont {A.}~\bibnamefont {Generalov}}, \bibinfo {author}
  {\bibfnamefont {K.}~\bibnamefont {Kliemt}}, \bibinfo {author} {\bibfnamefont
  {A.}~\bibnamefont {Kraiker}}, \bibinfo {author} {\bibfnamefont
  {C.}~\bibnamefont {Krellner}}, \bibinfo {author} {\bibfnamefont
  {K.}~\bibnamefont {Kummer}}, \bibinfo {author} {\bibfnamefont
  {S.}~\bibnamefont {Danzenb\"{a}cher}}, \bibinfo {author} {\bibfnamefont
  {C.}~\bibnamefont {Laubschat}}, \bibinfo {author} {\bibfnamefont {A.~P.}\
  \bibnamefont {Weber}}, \bibinfo {author} {\bibfnamefont {E.~V.}\ \bibnamefont
  {Chulkov}}, \bibinfo {author} {\bibfnamefont {A.~F.}\ \bibnamefont
  {Santander-Syro}}, \bibinfo {author} {\bibfnamefont {T.}~\bibnamefont
  {Imai}}, \bibinfo {author} {\bibfnamefont {K.}~\bibnamefont {Miyamoto}},
  \bibinfo {author} {\bibfnamefont {T.}~\bibnamefont {Okuda}},\ and\ \bibinfo
  {author} {\bibfnamefont {D.~V.}\ \bibnamefont {Vyalikh}},\ }\bibfield
  {title} {\bibinfo {title} {{Observation of a cubic Rashba effect in the
  surface spin structure of rare-earth ternary materials}},\ }\href
  {https://arxiv.org/abs/2002.01701} {\bibfield  {journal} {\bibinfo  {journal}
  {arXiv}\ ,\ \bibinfo {pages} {2002.01701}} (\bibinfo {year}
  {2020})}\BibitemShut {NoStop}%
\bibitem [{\citenamefont {Niesner}\ \emph {et~al.}(2012)\citenamefont
  {Niesner}, \citenamefont {Fauster}, \citenamefont {Eremeev}, \citenamefont
  {Menshchikova}, \citenamefont {Koroteev}, \citenamefont {Protogenov},
  \citenamefont {Chulkov}, \citenamefont {Tereshchenko}, \citenamefont {Kokh},
  \citenamefont {Alekperov}, \citenamefont {Nadjafov},\ and\ \citenamefont
  {Mamedov}}]{Niesner_PRB_2012}%
  \BibitemOpen
  \bibfield  {author} {\bibinfo {author} {\bibfnamefont {D.}~\bibnamefont
  {Niesner}}, \bibinfo {author} {\bibfnamefont {T.}~\bibnamefont {Fauster}},
  \bibinfo {author} {\bibfnamefont {S.~V.}\ \bibnamefont {Eremeev}}, \bibinfo
  {author} {\bibfnamefont {T.~V.}\ \bibnamefont {Menshchikova}}, \bibinfo
  {author} {\bibfnamefont {Y.~M.}\ \bibnamefont {Koroteev}}, \bibinfo {author}
  {\bibfnamefont {A.~P.}\ \bibnamefont {Protogenov}}, \bibinfo {author}
  {\bibfnamefont {E.~V.}\ \bibnamefont {Chulkov}}, \bibinfo {author}
  {\bibfnamefont {O.~E.}\ \bibnamefont {Tereshchenko}}, \bibinfo {author}
  {\bibfnamefont {K.~A.}\ \bibnamefont {Kokh}}, \bibinfo {author}
  {\bibfnamefont {O.}~\bibnamefont {Alekperov}}, \bibinfo {author}
  {\bibfnamefont {A.}~\bibnamefont {Nadjafov}},\ and\ \bibinfo {author}
  {\bibfnamefont {N.}~\bibnamefont {Mamedov}},\ }\bibfield  {title} {\bibinfo
  {title} {Unoccupied topological states on bismuth chalcogenides},\ }\href
  {https://doi.org/10.1103/PhysRevB.86.205403} {\bibfield  {journal} {\bibinfo
  {journal} {Phys. Rev. B}\ }\textbf {\bibinfo {volume} {86}},\ \bibinfo
  {pages} {205403} (\bibinfo {year} {2012})}\BibitemShut {NoStop}%
\bibitem [{\citenamefont {Sobota}\ \emph {et~al.}(2013)\citenamefont {Sobota},
  \citenamefont {Yang}, \citenamefont {Kemper}, \citenamefont {Lee},
  \citenamefont {Schmitt}, \citenamefont {Li}, \citenamefont {Moore},
  \citenamefont {Analytis}, \citenamefont {Fisher}, \citenamefont {Kirchmann},
  \citenamefont {Devereaux},\ and\ \citenamefont {Shen}}]{Sobota_PRL_2013}%
  \BibitemOpen
  \bibfield  {author} {\bibinfo {author} {\bibfnamefont {J.~A.}\ \bibnamefont
  {Sobota}}, \bibinfo {author} {\bibfnamefont {S.-L.}\ \bibnamefont {Yang}},
  \bibinfo {author} {\bibfnamefont {A.~F.}\ \bibnamefont {Kemper}}, \bibinfo
  {author} {\bibfnamefont {J.~J.}\ \bibnamefont {Lee}}, \bibinfo {author}
  {\bibfnamefont {F.~T.}\ \bibnamefont {Schmitt}}, \bibinfo {author}
  {\bibfnamefont {W.}~\bibnamefont {Li}}, \bibinfo {author} {\bibfnamefont
  {R.~G.}\ \bibnamefont {Moore}}, \bibinfo {author} {\bibfnamefont {J.~G.}\
  \bibnamefont {Analytis}}, \bibinfo {author} {\bibfnamefont {I.~R.}\
  \bibnamefont {Fisher}}, \bibinfo {author} {\bibfnamefont {P.~S.}\
  \bibnamefont {Kirchmann}}, \bibinfo {author} {\bibfnamefont {T.~P.}\
  \bibnamefont {Devereaux}},\ and\ \bibinfo {author} {\bibfnamefont {Z.-X.}\
  \bibnamefont {Shen}},\ }\bibfield  {title} {\bibinfo {title} {{Direct Optical
  Coupling to an Unoccupied Dirac Surface State in the Topological Insulator
  ${\mathrm{Bi}}_{2}{\mathrm{Se}}_{3}$}},\ }\href
  {https://doi.org/10.1103/PhysRevLett.111.136802} {\bibfield  {journal}
  {\bibinfo  {journal} {Phys. Rev. Lett.}\ }\textbf {\bibinfo {volume} {111}},\
  \bibinfo {pages} {136802} (\bibinfo {year} {2013})}\BibitemShut {NoStop}%
\bibitem [{\citenamefont {Niesner}\ \emph {et~al.}(2014)\citenamefont
  {Niesner}, \citenamefont {Otto}, \citenamefont {Fauster}, \citenamefont
  {Chulkov}, \citenamefont {Eremeev}, \citenamefont {Tereshchenko},\ and\
  \citenamefont {Kokh}}]{Niesner_JESRP_2014}%
  \BibitemOpen
  \bibfield  {author} {\bibinfo {author} {\bibfnamefont {D.}~\bibnamefont
  {Niesner}}, \bibinfo {author} {\bibfnamefont {S.}~\bibnamefont {Otto}},
  \bibinfo {author} {\bibfnamefont {T.}~\bibnamefont {Fauster}}, \bibinfo
  {author} {\bibfnamefont {E.~V.}\ \bibnamefont {Chulkov}}, \bibinfo {author}
  {\bibfnamefont {S.~V.}\ \bibnamefont {Eremeev}}, \bibinfo {author}
  {\bibfnamefont {O.~E.}\ \bibnamefont {Tereshchenko}},\ and\ \bibinfo {author}
  {\bibfnamefont {K.~A.}\ \bibnamefont {Kokh}},\ }\bibfield  {title} {\bibinfo
  {title} {Electron dynamics of unoccupied states in topological insulators},\
  }\href {https://doi.org/https://doi.org/10.1016/j.elspec.2014.03.013}
  {\bibfield  {journal} {\bibinfo  {journal} {Journal of Electron Spectroscopy
  and Related Phenomena}\ }\textbf {\bibinfo {volume} {195}},\ \bibinfo {pages}
  {258 } (\bibinfo {year} {2014})}\BibitemShut {NoStop}%
\bibitem [{\citenamefont {Datzer}\ \emph {et~al.}(2017)\citenamefont {Datzer},
  \citenamefont {Zumb\"ulte}, \citenamefont {Braun}, \citenamefont {F\"orster},
  \citenamefont {Schmidt}, \citenamefont {Mi}, \citenamefont {Iversen},
  \citenamefont {Hofmann}, \citenamefont {Min\'ar}, \citenamefont {Ebert},
  \citenamefont {Kr\"uger}, \citenamefont {Rohlfing},\ and\ \citenamefont
  {Donath}}]{Datzer_PRB_2017}%
  \BibitemOpen
  \bibfield  {author} {\bibinfo {author} {\bibfnamefont {C.}~\bibnamefont
  {Datzer}}, \bibinfo {author} {\bibfnamefont {A.}~\bibnamefont {Zumb\"ulte}},
  \bibinfo {author} {\bibfnamefont {J.}~\bibnamefont {Braun}}, \bibinfo
  {author} {\bibfnamefont {T.}~\bibnamefont {F\"orster}}, \bibinfo {author}
  {\bibfnamefont {A.~B.}\ \bibnamefont {Schmidt}}, \bibinfo {author}
  {\bibfnamefont {J.}~\bibnamefont {Mi}}, \bibinfo {author} {\bibfnamefont
  {B.}~\bibnamefont {Iversen}}, \bibinfo {author} {\bibfnamefont
  {P.}~\bibnamefont {Hofmann}}, \bibinfo {author} {\bibfnamefont
  {J.}~\bibnamefont {Min\'ar}}, \bibinfo {author} {\bibfnamefont
  {H.}~\bibnamefont {Ebert}}, \bibinfo {author} {\bibfnamefont
  {P.}~\bibnamefont {Kr\"uger}}, \bibinfo {author} {\bibfnamefont
  {M.}~\bibnamefont {Rohlfing}},\ and\ \bibinfo {author} {\bibfnamefont
  {M.}~\bibnamefont {Donath}},\ }\bibfield  {title} {\bibinfo {title}
  {{Unraveling the spin structure of unoccupied states in
  ${\mathrm{Bi}}_{2}{\mathrm{Se}}_{3}$}},\ }\href
  {https://doi.org/10.1103/PhysRevB.95.115401} {\bibfield  {journal} {\bibinfo
  {journal} {Phys. Rev. B}\ }\textbf {\bibinfo {volume} {95}},\ \bibinfo
  {pages} {115401} (\bibinfo {year} {2017})}\BibitemShut {NoStop}%
\bibitem [{\citenamefont {Aguilera}\ \emph {et~al.}(2019)\citenamefont
  {Aguilera}, \citenamefont {Friedrich},\ and\ \citenamefont
  {Bl\"ugel}}]{Aguilera_PRB_2019}%
  \BibitemOpen
  \bibfield  {author} {\bibinfo {author} {\bibfnamefont {I.}~\bibnamefont
  {Aguilera}}, \bibinfo {author} {\bibfnamefont {C.}~\bibnamefont
  {Friedrich}},\ and\ \bibinfo {author} {\bibfnamefont {S.}~\bibnamefont
  {Bl\"ugel}},\ }\bibfield  {title} {\bibinfo {title} {Many-body corrected
  tight-binding hamiltonians for an accurate quasiparticle description of
  topological insulators of the ${\mathrm{bi}}_{2}{\mathrm{se}}_{3}$ family},\
  }\href {https://doi.org/10.1103/PhysRevB.100.155147} {\bibfield  {journal}
  {\bibinfo  {journal} {Phys. Rev. B}\ }\textbf {\bibinfo {volume} {100}},\
  \bibinfo {pages} {155147} (\bibinfo {year} {2019})}\BibitemShut {NoStop}%
\bibitem [{\citenamefont {Krasovskii}(1997)}]{Krasovskii_PRB_1997}%
  \BibitemOpen
  \bibfield  {author} {\bibinfo {author} {\bibfnamefont {E.~E.}\ \bibnamefont
  {Krasovskii}},\ }\bibfield  {title} {\bibinfo {title} {Accuracy and
  convergence properties of the extended linear augmented-plane-wave method},\
  }\href {https://doi.org/10.1103/PhysRevB.56.12866} {\bibfield  {journal}
  {\bibinfo  {journal} {Phys. Rev. B}\ }\textbf {\bibinfo {volume} {56}},\
  \bibinfo {pages} {12866} (\bibinfo {year} {1997})}\BibitemShut {NoStop}%
\bibitem [{\citenamefont {Krasovskii}\ \emph {et~al.}(1999)\citenamefont
  {Krasovskii}, \citenamefont {Starrost},\ and\ \citenamefont
  {Schattke}}]{Krasovskii_PRB_1999}%
  \BibitemOpen
  \bibfield  {author} {\bibinfo {author} {\bibfnamefont {E.~E.}\ \bibnamefont
  {Krasovskii}}, \bibinfo {author} {\bibfnamefont {F.}~\bibnamefont
  {Starrost}},\ and\ \bibinfo {author} {\bibfnamefont {W.}~\bibnamefont
  {Schattke}},\ }\bibfield  {title} {\bibinfo {title} {Augmented fourier
  components method for constructing the crystal potential in self-consistent
  band-structure calculations},\ }\href
  {https://doi.org/10.1103/PhysRevB.59.10504} {\bibfield  {journal} {\bibinfo
  {journal} {Phys. Rev. B}\ }\textbf {\bibinfo {volume} {59}},\ \bibinfo
  {pages} {10504} (\bibinfo {year} {1999})}\BibitemShut {NoStop}%
\bibitem [{\citenamefont {Koelling}\ and\ \citenamefont
  {Harmon}(1977)}]{Koelling_1977}%
  \BibitemOpen
  \bibfield  {author} {\bibinfo {author} {\bibfnamefont {D.~D.}\ \bibnamefont
  {Koelling}}\ and\ \bibinfo {author} {\bibfnamefont {B.~N.}\ \bibnamefont
  {Harmon}},\ }\bibfield  {title} {\bibinfo {title} {A technique for
  relativistic spin-polarised calculations},\ }\href
  {https://doi.org/10.1088/0022-3719/10/16/019} {\bibfield  {journal} {\bibinfo
   {journal} {Journal of Physics C: Solid State Physics}\ }\textbf {\bibinfo
  {volume} {10}},\ \bibinfo {pages} {3107} (\bibinfo {year}
  {1977})}\BibitemShut {NoStop}%
\bibitem [{\citenamefont {Wyckoff}(1964)}]{Wyckoff_RWG}%
  \BibitemOpen
  \bibfield  {author} {\bibinfo {author} {\bibfnamefont {R.~W.~G.}\
  \bibnamefont {Wyckoff}},\ }\href@noop {} {\emph {\bibinfo {title} {{Crystal
  Structures 2}}}}\ (\bibinfo  {publisher} {John Wiley and Sons},\ \bibinfo
  {address} {New York},\ \bibinfo {year} {1964})\BibitemShut {NoStop}%
\bibitem [{\citenamefont {Nechaev}\ \emph {et~al.}(2013)\citenamefont
  {Nechaev}, \citenamefont {Hatch}, \citenamefont {Bianchi}, \citenamefont
  {Guan}, \citenamefont {Friedrich}, \citenamefont {Aguilera}, \citenamefont
  {Mi}, \citenamefont {Iversen}, \citenamefont {Bl\"ugel}, \citenamefont
  {Hofmann},\ and\ \citenamefont {Chulkov}}]{Nechaev_PRB_2013_BISE}%
  \BibitemOpen
  \bibfield  {author} {\bibinfo {author} {\bibfnamefont {I.~A.}\ \bibnamefont
  {Nechaev}}, \bibinfo {author} {\bibfnamefont {R.~C.}\ \bibnamefont {Hatch}},
  \bibinfo {author} {\bibfnamefont {M.}~\bibnamefont {Bianchi}}, \bibinfo
  {author} {\bibfnamefont {D.}~\bibnamefont {Guan}}, \bibinfo {author}
  {\bibfnamefont {C.}~\bibnamefont {Friedrich}}, \bibinfo {author}
  {\bibfnamefont {I.}~\bibnamefont {Aguilera}}, \bibinfo {author}
  {\bibfnamefont {J.~L.}\ \bibnamefont {Mi}}, \bibinfo {author} {\bibfnamefont
  {B.~B.}\ \bibnamefont {Iversen}}, \bibinfo {author} {\bibfnamefont
  {S.}~\bibnamefont {Bl\"ugel}}, \bibinfo {author} {\bibfnamefont
  {P.}~\bibnamefont {Hofmann}},\ and\ \bibinfo {author} {\bibfnamefont {E.~V.}\
  \bibnamefont {Chulkov}},\ }\bibfield  {title} {\bibinfo {title} {{Evidence
  for a direct band gap in the topological insulator Bi${}_{2}$Se${}_{3}$ from
  theory and experiment}},\ }\href {https://doi.org/10.1103/PhysRevB.87.121111}
  {\bibfield  {journal} {\bibinfo  {journal} {Phys. Rev. B}\ }\textbf {\bibinfo
  {volume} {87}},\ \bibinfo {pages} {121111} (\bibinfo {year}
  {2013})}\BibitemShut {NoStop}%
\bibitem [{\citenamefont {L\"{o}wdin}(1951)}]{Leowdin_JCP_1951}%
  \BibitemOpen
  \bibfield  {author} {\bibinfo {author} {\bibfnamefont {P.-O.}\ \bibnamefont
  {L\"{o}wdin}},\ }\bibfield  {title} {\bibinfo {title} {{A Note on the
  Quantum-Mechanical Perturbation Theory}},\ }\href
  {https://doi.org/10.1063/1.1748067} {\bibfield  {journal} {\bibinfo
  {journal} {The Journal of Chemical Physics}\ }\textbf {\bibinfo {volume}
  {19}},\ \bibinfo {pages} {1396} (\bibinfo {year} {1951})}\BibitemShut
  {NoStop}%
\bibitem [{\citenamefont {Schrieffer}\ and\ \citenamefont
  {Wolff}(1966)}]{Schrieffer_PR_1966}%
  \BibitemOpen
  \bibfield  {author} {\bibinfo {author} {\bibfnamefont {J.~R.}\ \bibnamefont
  {Schrieffer}}\ and\ \bibinfo {author} {\bibfnamefont {P.~A.}\ \bibnamefont
  {Wolff}},\ }\bibfield  {title} {\bibinfo {title} {{Relation between the
  Anderson and Kondo Hamiltonians}},\ }\href
  {https://doi.org/10.1103/PhysRev.149.491} {\bibfield  {journal} {\bibinfo
  {journal} {Phys. Rev.}\ }\textbf {\bibinfo {volume} {149}},\ \bibinfo {pages}
  {491} (\bibinfo {year} {1966})}\BibitemShut {NoStop}%
\bibitem [{\citenamefont {Winkler}(2003)}]{Winkler_KP}%
  \BibitemOpen
  \bibfield  {author} {\bibinfo {author} {\bibfnamefont {R.}~\bibnamefont
  {Winkler}},\ }\href@noop {} {\emph {\bibinfo {title} {{Spin-Orbit Coupling
  Effects in Two-Dimensional Electron and Hole Systems}}}}\ (\bibinfo
  {publisher} {Springer},\ \bibinfo {address} {Berlin},\ \bibinfo {year}
  {2003})\BibitemShut {NoStop}%
\bibitem [{\citenamefont {Fu}(2009)}]{Fu_PRL_2009}%
  \BibitemOpen
  \bibfield  {author} {\bibinfo {author} {\bibfnamefont {L.}~\bibnamefont
  {Fu}},\ }\bibfield  {title} {\bibinfo {title} {{Hexagonal Warping Effects in
  the Surface States of the Topological Insulator
  ${\mathrm{Bi}}_{2}{\mathrm{Te}}_{3}$}},\ }\href
  {https://doi.org/10.1103/PhysRevLett.103.266801} {\bibfield  {journal}
  {\bibinfo  {journal} {Phys. Rev. Lett.}\ }\textbf {\bibinfo {volume} {103}},\
  \bibinfo {pages} {266801} (\bibinfo {year} {2009})}\BibitemShut {NoStop}%
\bibitem [{Note1()}]{Note1}%
  \BibitemOpen
  \bibinfo {note} {{A multiple winding of the in-plane spin was recently
  observed at the Si-terminated surface of TbRh$_2$Si$_2$ (the $C_{4v}$ crystal
  symmetry)~[\onlinecite {Usachov_arXiv_2020}]. The four-fold symmetric field
  ${\protect \bm {\protect \mathcal {B}}}^{(3)} = k^3 (\protect \qopname \relax
  o{sin}3\varphi _{\protect \mathbf {k}}, \protect \qopname \relax
  o{cos}3\varphi _{\protect \mathbf {k}}, 0)$ was proved to cause a triple
  winding of the surface-state spin. Here, the non-orthogonality is $\protect
  \qopname \relax o{sin}\delta ^{\pm } = \pm \protect \tilde {\gamma
  }k^3\protect \qopname \relax o{sin}4\varphi _{\protect \mathbf {k}}/|\protect
  \bm {\protect \mathcal {B}}_{\shortparallel }|$, with the in-plane field
  $|\protect \bm {\protect \mathcal {B}}_{\shortparallel }|^2 = (\protect
  \tilde {\alpha }^2+\protect \tilde {\gamma }^2k^4)k^2-2\protect \tilde
  {\alpha } \protect \tilde {\gamma }k^4\protect \qopname \relax o{cos}4\varphi
  _{\protect \mathbf {k}}$ in the notation of Ref.~[\onlinecite {Usachov_arXiv_2020}].
  The cubic in-plane field contributes to the four-fold warping of the CECs
  through the scalar product ${\protect \bm {\protect \mathcal {B}}}^{(3)}\cdot
  {\protect \bm {\protect \mathcal {B}}}_{\protect \mathrm {R}}^{(1)}$, as in
  the present study. Note that an additional contribution to both the locking
  angle and the warping may also come from the four-fold symmetric field
  ${\setbox \z@ \hbox {\mathsurround \z@ $\textstyle \protect \bm {\protect
  \mathcal {B}}$}\mathaccent "0365{\protect \bm {\protect \mathcal {B}}}}^{(5)}
  = k^5 (\protect \qopname \relax o{sin}5\varphi _{\protect \mathbf {k}},
  -\protect \qopname \relax o{cos}5\varphi _{\protect \mathbf {k}},
  0)$.}}\BibitemShut {Stop}%
\bibitem [{\citenamefont {Krasovskii}(2014)}]{Krasovskii_PRB_2014}%
  \BibitemOpen
  \bibfield  {author} {\bibinfo {author} {\bibfnamefont {E.~E.}\ \bibnamefont
  {Krasovskii}},\ }\bibfield  {title} {\bibinfo {title} {Microscopic origin of
  the relativistic splitting of surface states},\ }\href
  {https://doi.org/10.1103/PhysRevB.90.115434} {\bibfield  {journal} {\bibinfo
  {journal} {Phys. Rev. B}\ }\textbf {\bibinfo {volume} {90}},\ \bibinfo
  {pages} {115434} (\bibinfo {year} {2014})}\BibitemShut {NoStop}%
\end{thebibliography}
\end{document}